\documentclass{nature-pre}
\usepackage{times}
\usepackage{amsmath}
\usepackage{amssymb}
\usepackage{physics}	
\usepackage{comment}
\usepackage{multirow}
\usepackage{physics}
\usepackage{makecell}

\usepackage[utf8]{inputenc}
\usepackage{amsmath}
\usepackage{amssymb}
\usepackage{physics}	
\usepackage{comment}
\usepackage{multirow}
\usepackage{siunitx}

\usepackage{graphicx}
\graphicspath{./}
\usepackage{placeins}
\usepackage{threeparttable}
\usepackage{float}
\usepackage{placeins}
	
\usepackage{color, colortbl}
\definecolor{lightgray}{gray}{0.9}

 \usepackage[usenames,dvipsnames]{xcolor}
 \usepackage[normalem]{ulem}
 \usepackage[version=4]{mhchem}
\usepackage{xr-hyper}
\makeatletter
\newcommand*{\addFileDependency}[1]{
\typeout{(#1)}
\@addtofilelist{#1}
\IfFileExists{#1}{}{\typeout{No file #1.}}
}\makeatother

\newcommand*{\myexternaldocument}[1]{%
\externaldocument{#1}%
\addFileDependency{#1.tex}%
\addFileDependency{#1.aux}%
}
\myexternaldocument{REVISED3-SupplementaryInfo}

\usepackage{hyperref}
\hypersetup{
 bookmarks=true,		
 unicode=false,			
 pdftoolbar=true,		
 pdffitwindow=false,		
 pdfstartview={FitH},		
 pdfcreator={pdflatex},		
 pdfproducer={vim},		
 pdfnewwindow=true,		
 colorlinks=true,		
 linktoc=page,			
 linkcolor=blue,		
 citecolor=blue,		
 filecolor=blue,		
 urlcolor=blue			
}

\newcommand{\VBm}{V$_{\text{B}}^-$}

\usepackage{array}

\title{Accurate description of the electronic states of the \VBm{} center in hBN by wavefunction theory}

\author{Zsolt Benedek$^{1,2,*}$, \'{A}d\'{a}m Ganyecz$^{2,3}$, Oscar Bulancea-Lindvall$^{4}$,  Gergely Barcza$^{2,3,*}$, Viktor Iv\'{a}dy$^{1,2,*}$}

\begin{document}
\maketitle

\begin{affiliations}
\item{Department of Physics of Complex Systems, Eötvös Loránd University, Egyetem tér 1-3, H-1053 Budapest, Hungary}
\item {MTA-ELTE Lend\"{u}let "Momentum" NewQubit Research Group, Pázmány Péter, Sétány 1/A, 1117 Budapest, Hungary}
\item  {HUN-REN Wigner Research Centre for Physics, PO Box 49, H-1525, Budapest, Hungary}
\item{Department of Physics, Chemistry and Biology (IFM), Link\"oping University, SE-581 83 Link\"oping, Sweden}
\item[*] email: benedek.zsolt@ttk.elte.hu, barcza.gergely@wigner.hu, ivady.viktor@ttk.elte.hu
\end{affiliations}

\date{\today}

\begin{abstract}

\vspace{1cm}

{\large Abstract}

\vspace{1cm}


Ensembles of negatively charged boron vacancy (\VBm) centers in hexagonal boron nitride (hBN) have emerged as a two-dimensional spin qubit system interfaced with optics to advance nanoscale quantum sensing. However, a comprehensive description of its optically detected magnetic resonance (ODMR) signal remains challenging due to the strongly correlated nature of the excited electronic states involved in its optical cycle. In this work, we model the energetics, structural relaxation, and transition rates of the \VBm{} center using a high-level wave-function-based electron correlation method (CASSCF-NEVPT2). We provide a thorough analysis of the excited state fine structure and pseudo Jahn-Teller effects, singlet-triplet quasi-degeneracies, photoluminescence parameters, intersystem crossing pathways, and stress-dependence of the fine structure and decay parameters. Our findings not only clarify the fundamental behavior of the \VBm{} center in hBN but also establish the theoretical foundation for advancing the \VBm{} center's readout for integrated 2D quantum sensors.

\end{abstract}

\maketitle

\newpage

\section*{\large Introduction}

Color centers in solids created by point defects are becoming essential components in emerging quantum applications~\cite{weber_quantum_2010,wolfowicz_quantum_2021, Kianinia_2022,esmann2024}. Point defects with spin-dependent photoluminescence (PL) enabling an optically detected magnetic resonance (ODMR) signal can implement an optically addressable quantum bit operating under ambient conditions. Historically, the negatively charged nitrogen-vacancy (NV) center in diamond~\cite{Doherty2011} has been the gold standard for such applications~\cite{rondin_magnetometry_2014,rovny_nanoscale_2024}. However, traditional three-dimensional host materials face a fundamental physical limitation, namely, achieving nanoscopic spatial resolution requires positioning the defects extremely close to the host's surface, which inherently leads to severe surface-induced instability and rapid spin decoherence~\cite{romach_spectroscopy_2015,sangtawesin_origins_2019}.

To overcome this barrier, two-dimensional (2D) van der Waals materials have recently emerged as host materials for next-generation optically addressable spin qubits. Among these, hexagonal boron nitride (hBN) stands out due to its wide bandgap, exceptional chemical stability, and ability to host ultra-bright emitters~\cite{tran_quantum_2016} and room-temperature ODMR centers~\cite{gottscholl_2020,chejanovsky_single-spin_2021,stern_room-temperature_2022,gao_single_2025,robertson_charge_2025,whitefield_narrowband_2026}. Because the surface of hBN can be atomically flat and the thickness of the host can be engineered by exfoliation, defects in hBN can reside in close proximity to the target, allowing deterministic engineering of sensor-sample distances~\cite{tetienne_quantum_2021,healey_quantum_2022}, which can be considerably smaller than those achievable in bulk host materials such as diamond.

A particularly promising defect in hBN is the negatively charged boron vacancy (\VBm) center~\cite{gottscholl_2020,ivady_ab_2020}. It can be created in large concentrations by thermal neutron irradiation, features an $S=1$ ground state with a well-known ODMR spectrum at room temperature~\cite{gottscholl_initialization_2020,gottscholl_2020,gottscholl_spin_2021}, and is susceptible to strain, temperature, electric, and magnetic fields~\cite{gottscholl_spin_2021,gao_high-contrast_2021,lyu_strain_2022}. These attributes make the \VBm{} center a prime candidate for nanoscale quantum sensing~\cite{daly_prospects_2025,biswas_quantum_2025}.

Despite these recent promising results, there is still room for improvement. Understanding the electronic structure and the optical excitation and decay pathways is crucial for devising a strong ODMR signal for readout. First-principles calculations play a critical role here as they provide deep insight into microscopic processes beyond the capabilities of experimental methods.

Early theoretical works focused on the charge transition levels of the defect center~\cite{Attaccalite2011,Tawfik2017,Abdi2018,weston_native_2018}. More recently, details of the energy spectrum have been revealed by density functional theory and wave-function theory approaches~\cite{Tawfik2017,Abdi2018,ivady_ab_2020,sajid_edge_2020,reimers_photoluminescence_2020,Barcza_2021}. Currently, the ODMR signal of the $D_{3h}$-symmetric $^3\bar{\text{A}}'_2$ ground state is hypothesized to be initiated by optical excitation to the $^3\bar{\text{E}}''$ state, followed by a Jahn-Teller distortion to $C_{2v}$ symmetry, resulting in the $^3\bar{\text{A}}_2$ excited state. This state may decay not only optically but also through a spin-selective intersystem crossing to the $^1\bar{\text{E}}' \rightarrow {}^1\bar{\text{A}}_1$ singlet shelving state~\cite{ivady_ab_2020,reimers_photoluminescence_2020}.

However, the complete ODMR process of the \VBm{} center remains incomprehended owing to the highly correlated, multireference nature of the involved electronic states~\cite{ivady_ab_2020, Reimers2018, reimers_photoluminescence_2020}. In this article, we provide an extended theory of the ODMR properties of \VBm{} using state-of-the-art \textit{ab initio} modeling. Employing complete active space self-consistent field (CASSCF) methods augmented by $n$-electron valence perturbation theory (NEVPT2), we detail the excitation energies, structural relaxation effects, and transition rates. We reveal higher-lying states, explore the static and intermediate static-dynamic Jahn-Teller nature of the excited state, identify distinct pathways taken by the $m_S = 0$ and $m_S = \pm1$ states, and propose possible two-laser excitation pathways. Finally, we investigate relevant measurable quantities under external strain, providing crucial theoretical support for the deployment of \VBm{} centers as integrated 2D quantum sensors.

\section*{\large Results}
\label{sec:results}


We begin our discussion by presenting the energy spectrum of the \VBm{} center and comparing it with available theoretical and experimental data. In contrast to previous theoretical studies, here we employ a completely wavefunction theory (WFT) based approach,  which has proven to be successful in benchmarking the excited states of the prototypical NV$^-$ center in diamond~\cite{benedek_accurate_2025,luu_identifying_2025}, to obtain the electronic structure and optimal geometry. For the former, we employ the CASSCF-NEVPT2 method as our highest-level approximation to the Schrödinger equation, which describes both static and dynamical correlation effects. For the latter, we employ the CASSCF method. For more details on the applied methodologies, we refer to Methods and Supplementary Notes~1-2.

\textcolor{black}{As WFT methods are currently only available for cluster models in conventional quantum chemistry codes, we perform the calculations for 5 cluster models with increasing number of boron-nitrogen shells around the \VBm{} defect, investigating the convergence with respect to model size from MODEL-1 (\ce{B6N6H12-}) to MODEL-5 (\ce{B90N90H33-}) - see Supplementary Figure 1 for visual explanation. Unless noted otherwise, the results are presented herein for MODEL-4 (\ce{B60N60H27-}), while model size dependence is discussed under Supplementary Tables. } 

\begin{figure*}[!h]
    \centering
    \includegraphics[width=0.69\linewidth]{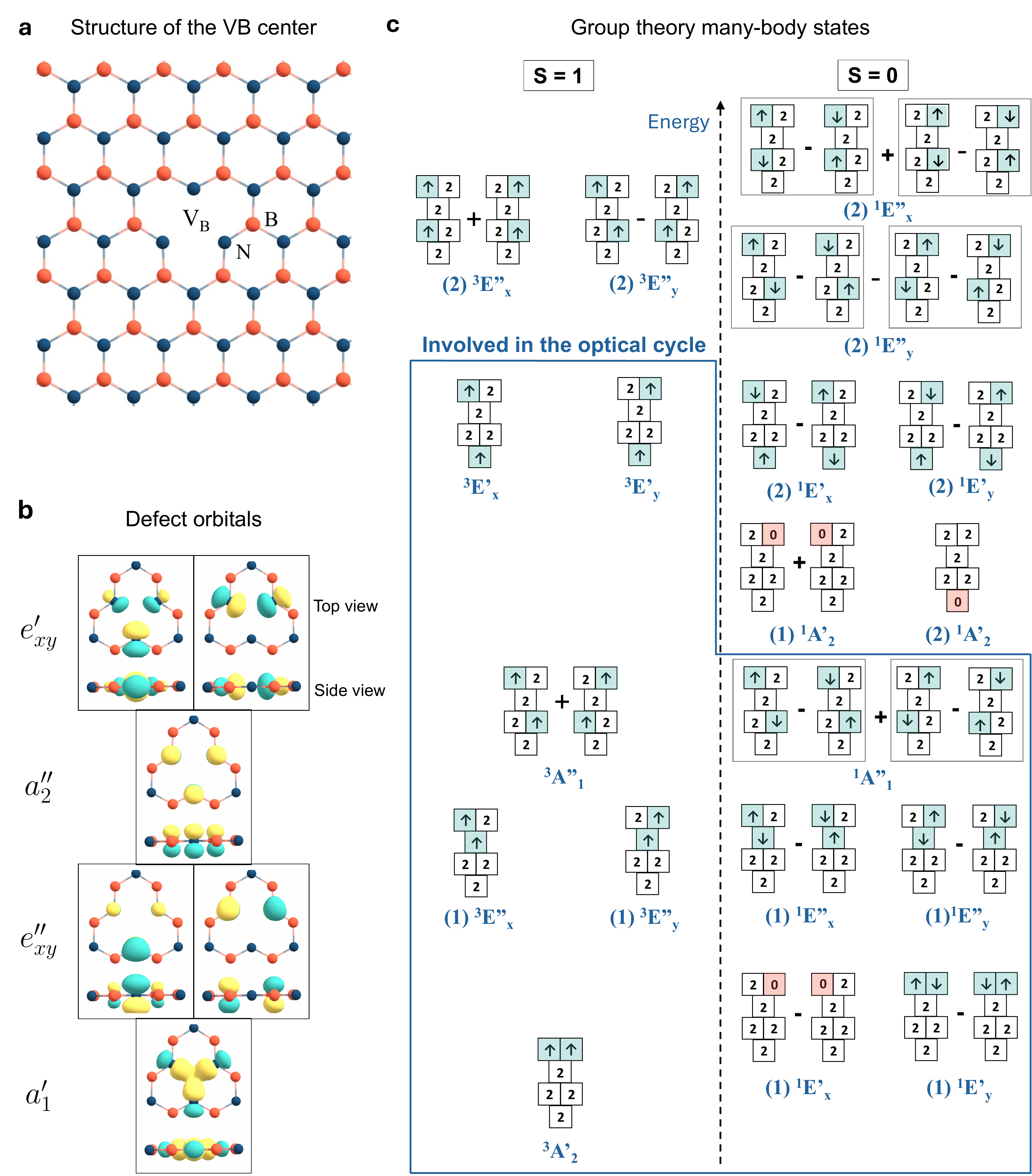}
    \caption{ Structure and electronic spectrum of the  \VBm{}.  \textbf{a} Structure of the \VBm{} center in hexagonal boron-nitride sheet. \textbf{b} Six defect orbitals of the \VBm{} center localized on the first neighbor nitrogen atoms. The states are labelled according to their group theory irreducible representation, reflecting the states' symmetry properties. Colored lobes around the nitrogen atoms are isosurfaces of the wavefunctions, with turquoise and yellow colors representing positive and negative values of the wavefunctions. \textbf{c} Symmetry-adapted combination of Slater determinants of the ground and excited states as derived from group theory. The numbers in the boxes indicate the occupancy of the orbitals in \textbf{b}. Colored boxes highlight partially filled (light teal) and empty (pink) states. There are single Slater determinant many-body states, such as the ${}^3A^{\prime}_2$ ground state, and multideterminant states, such as the ${}^1E^\prime$.
    }
    \label{fig:fig1}
\end{figure*}

\begin{table}[h]
\setlength\extrarowheight{4pt}
\centering

\caption{Composition of the low-lying states of the \VBm{} center. The CASSCF many-body eigenstates (denoted with bars) are expanded in the basis of the symmetry adapted configurations provided in Fig.~\ref{fig:fig1}. For each state, the vertical CASSCF-NEVPT2 energy is also provided at ground-state atomic configuration. Due to the fixed geometry, these energy levels approximately correspond to the absorption maxima of the respective states. }

\resizebox{\textwidth}{!}{
\begin{tabular}{|l|l|c||l|l|c|}
\hline
\multicolumn{3}{|c||}{\textbf{\textcolor{black}{S = 1}}} & \multicolumn{3}{|c|}{\textbf{\textcolor{black}{S = 0}}}    \\
\hline 
State & \multicolumn{1}{c|}{Weights} & $E_\text{vertical}$ (eV) & State & \multicolumn{1}{c|}{Weights} & $E_\text{vertical}$ (eV) \\
\hline
 ${}^3\bar{A}_2^\prime$ & $\sim$100 \% ${}^3A_2^\prime$ & 0.00 & ${}^1\bar{E}^\prime_{x/y}$ & \phantom{$\sim$1}78 \% $(1){}^1E’_{x/y}$ + 22 \% $(2){}^1E^\prime_{x/y} $ & 0.78  \\
 \textcolor{black}{${}^3\bar{A}^{\prime\prime}_1$} & $\sim$100 \% \textcolor{black}{${}^3A^{\prime\prime}_1$} & 2.22 & \textcolor{black}{${}^1\bar{A}^{\prime\prime}_1$} & $\sim$100 \% \textcolor{black}{$(1){}^1A^{\prime\prime}_1$}  &2.19
 \\
${}^3\bar{E}^{\prime\prime}_{x/y}$ & \phantom{$\sim$1}46 \% $(1){}^3E^{\prime\prime}_{x/y}$  + 54 \% $(2){}^3E^{\prime\prime}_{x/y}$ & 2.24 & ${}^1\bar{E}^{\prime\prime}_{x/y}$ & \phantom{$\sim$1}46 \% $(1){}^1E^{\prime\prime}_{x/y}$ + 54 \% $(2){}^1E^{\prime\prime}_{x/y}$  & 2.25
  \\
${}^3\bar{E}^{\prime}_{x/y}$ & $\sim$100 \% $(1){}^3E^{\prime}_{x/y}$ & 2.83 & $ {}^1\bar{A}^{\prime}_1$ & \phantom{$\sim$1}64 \% $(1){}^1A^{\prime}_{1\phantom{/y}}$ + 36 \% $(2){}^1A^{\prime}_1$ & 2.75
 \\
\hline
\end{tabular}
}

\label{table:CASSCF}  
\end{table}

As detailed in previous works\cite{ivady_ab_2020,reimers_photoluminescence_2020}, the removal of the boron atom gives rise to 6 defect states, 3 in-plane dangling bonds forming an $a_1^{\prime}$ state and a degenerate $e^{\prime}$ state, and 3 out-of-plane $p_z$ orbitals forming an $a_2^{\prime\prime}$ state and a degenerate $e^{\prime\prime}$ state, all of which are localized on the first neighbor nitrogen atoms (Figure~\ref{fig:fig1}a,b). These six states form the basis for expanding the many-body states and define the active space for high-level wavefunction theory methods, where static correlation effects are taken into account. For comparison, recall that the essential active space of the NV center in diamond can be formed of only four defect states, i.e., two $a_1$ states and a degenerate $e$ state~\cite{benedek_accurate_2025}. Symmetry and the different correlated bases are the root cause of the deviations between the two systems, as detailed later in our work.


The wavefunctions of the conceivable electronic states of \VBm{} can be expressed in terms of symmetry-adapted configuration state functions. The leading configurations of low-lying man-body states, labeled by their corresponding irreducible representations, are provided in Fig.~\ref{fig:fig1},c. In our notation, the configurations derived from group theory\cite{ivady_ab_2020} are labeled as ${}^sX$, where $X \in \left\lbrace \text{A}_2^{\prime}, \textcolor{black}{\text{A}_1^{\prime\prime}}, \text{E}^{\prime}, \text{E}^{\prime\prime}  \right\rbrace$ and $ X \in \left\lbrace \text{A}_1, \text{A}_2, \text{B}_1, \text{B}_2 \right\rbrace$ in $D_{3h}$ and C$_{2v}$ symmetry,  respectively, and $s$ superscript provides the spin multiplicity of the state. Due to static correlation effects, the simple group theory picture is altered by state mixing. While group theory and single-reference DFT cannot provide information on the degree of mixing, such quantities can be directly obtained from many-body wavefunction theory. In the following, the many-body electronic states obtained by the CASSCF approach are indicated by the upper bar in our notation.  As an illustration,  in Table~\ref{table:CASSCF}, we express the composition of the low-lying CASSCF vertical eigenstates  at the ground-state geometry in the basis of  $D_{3h}$-symmetry-adapted configurations given in Fig.~\ref{fig:fig1},c. We note that symmetry reduction enables additional state mixing via the pseudo Jahn-Teller effect, which will be discussed later.  

\begin{figure*}[!h]
    \centering
    \includegraphics[width=0.90\linewidth]{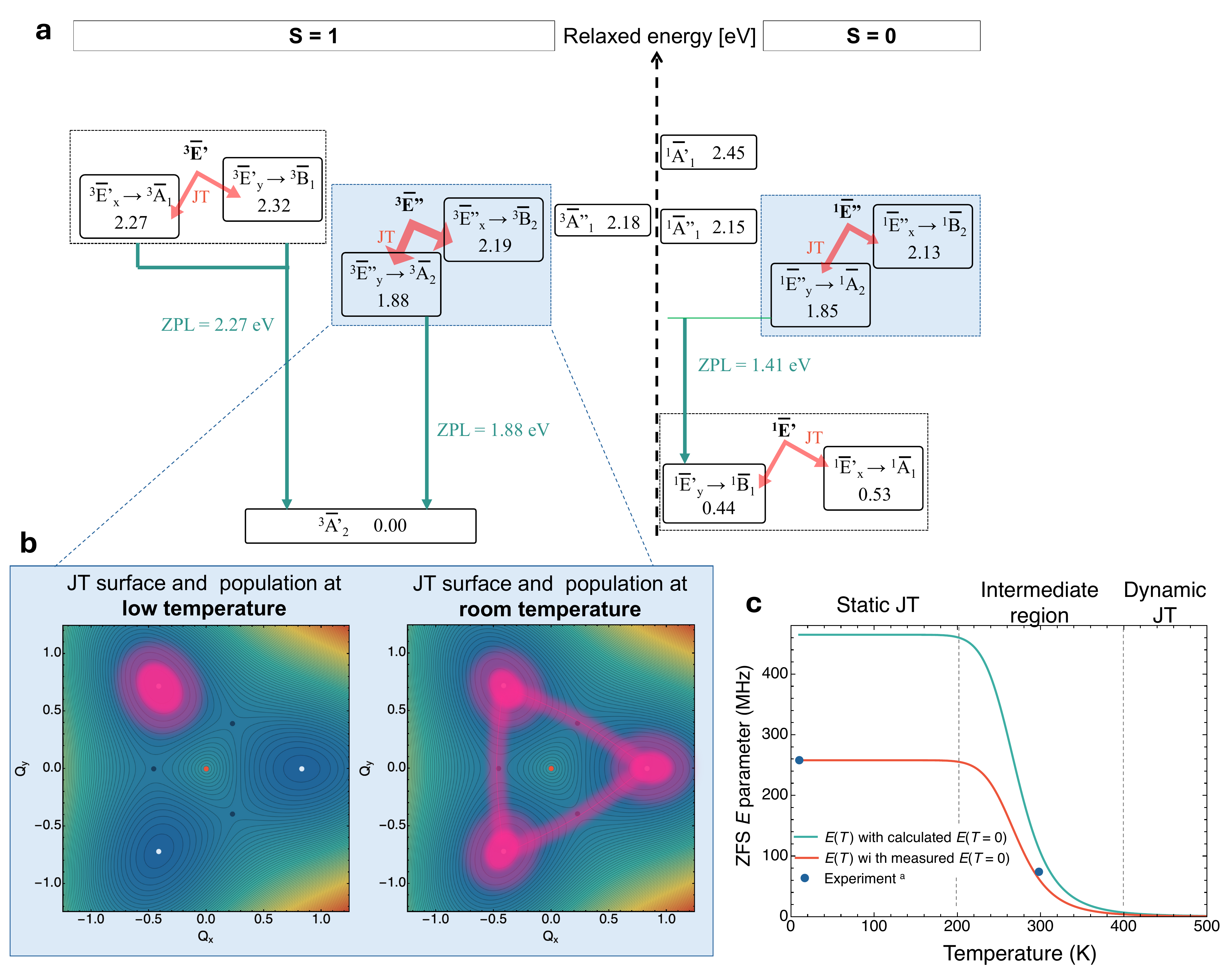}
    \caption{Spectrum and structural relaxation of the excited states. 
\textbf{a} Energy spectrum of the \VBm{} center obtained after excited-state relaxation. Boxes indicate Jahn-Teller unstable configurations and the energies of the resulting symmetry-breaking configurations. 
\textbf{b} \textcolor{black}{ Schematic illustration of the pseudo Jahn-Teller potential energy surface of the lowest triplet excited state in the plane of a representative pair of $e'_x$ and $e'_y$ symmetry-breaking displacements. The plot is not calculated data, as the Jahn-Teller distortion also involves the totally symmetric $a'_1$ mode and further $e'$ modes localized on the second-nearest neighbors, which cannot be faithfully represented in two dimensions.   Pink clouds indicate the probability density of the structure at low temperature (left), where the system remains in one of the three equivalent minima during the excited-state lifetime, and at high temperature (right), where thermally activated hopping between the minima averages the transverse zero-field splitting.} \textbf{c} Temperature dependence of the excited-state transverse zero-field splitting parameter ($E$). The teal and red curves are calculated using either the theoretical or experimental low-temperature values of the transverse zero-field splitting parameter. Blue points represent experimental measurements~\cite{Mathur2022} at low and high temperatures.
    }
    \label{fig:fig2}
\end{figure*}

After relaxing each electronic state at the many-body wavefunction theory level, we obtain an energy spectrum (Fig.~\ref{fig:fig2}a) which is qualitatively consistent with previous computational and theoretical results\cite{ivady_ab_2020,reimers_photoluminescence_2020}, but also provides a more detailed insight into the electronic structure of the \VBm{} center.
The first three excited states above the $^{3}\bar{\text{A}}_2^{\prime}$ ground state are the \textcolor{black}{$^{3}\bar{\text{A}}_1^{\prime\prime}$}, $^{3}\bar{\text{E}}^{\prime\prime}$, and $^{3}\bar{\text{E}}^{\prime}$ states in $D_{3h}$ symmetry, in increasing order of energy. 
As discussed previously\cite{ivady_ab_2020,reimers_photoluminescence_2020} and confirmed by our present calculations, the only optically allowed transition occurs between the ground state and the third (disregarding orbital multiplicities) $^{3}\bar{\text{E}}^{\prime}$ excited state. 
The calculated radiative transition rate, $1/\tau_{\text{rad},\,^{3}\bar{\text{E}}^{\prime}}$, is as high as 200~MHz, which enables efficient optical pumping of the \VBm{} center with $\approx$2.3 eV photon energy, i.e., a green laser (see Supplementary Table~9 for details). 

The optically excited \VBm{} center relaxes through rapid internal conversion (IC) and phonon emission into the lower-lying triplet \textcolor{black}{$^{3}\bar{\text{A}}_1^{\prime\prime}$} and $^{3}\bar{\text{E}}^{\prime\prime}$ states. 
While the former state retains nearly the same geometry as the ground state and preserves $D_{3h}$ symmetry, the $^{3}\bar{\text{E}}^{\prime\prime}$ state undergoes a substantial Jahn-Teller distortion, splitting the orbitally degenerate $^{3}\bar{\text{E}}^{\prime\prime}$ into a lower-lying $ {^{3}\bar{\text{A}}_2}$ state and a higher-lying $ {^{3}\bar{\text{B}}_2}$ state. (See Supplementary Note 3 for a detailed discussion on Jahn-Teller effects in \VBm{} .)
The $^{3}\bar{\text{E}}_y^{\prime\prime} \rightarrow {^{3}\bar{\text{A}}_2}$  relaxation reverses the energy ordering of the triplet excited states and the distorted ${^{3}\bar{\text{A}}_2}$ configuration becomes the lowest energy triplet excited state in C$_{2\text{v}}$ symmetry (as highlighted in Fig.~\ref{fig:fig2}a).
According to our calculations, the Jahn-Teller distortion leads to atomic mass-weighted displacements of $\Delta Q_{A_2} =$~0.83~$\sqrt{\text{amu}}$\,\AA\   relative to the high-symmetry configuration, with the corresponding Jahn-Teller stabilization energy of $E_{\text{JT}} = 355$~meV. 
The originally equilateral triangle of inner N atoms becomes stretched in one direction, possessing a short 2.49 \AA\  and two long 2.65 \AA\  N-N interatomic distances.
The pronounced structural deformation of the $^{3}\bar{\text{A}}_2$ state can be attributed to an intensive pseudo-Jahn-Teller effect between the reduced-symmetry $^{3}\bar{\text{E}}_y^{\prime\prime} \rightarrow {^{3}\bar{\text{A}}_2}$ and \textcolor{black}{$^{3}\bar{\text{A}}_1^{\prime\prime} \rightarrow {^{3}\bar{\text{A}}_2}$} states. This is due to fact that the mixed $^{3}\bar{\text{E}}_y^{\prime\prime}$ state and \textcolor{black}{$^{3}\bar{\text{A}}_1^{\prime\prime}$} states share two leading determinants and are close in energy (see Fig.~\ref{fig:fig1}c and Table 1), which leads to strong electron-phonon coupling for $e^{\prime}$-type vibrations. \textcolor{black}{The coupling between $^3\bar{B}_2$ and $^3A''_1$ is mediated solely by the symmetry-lowering $b_1$ component of $e'$ vibrational mode and is therefore absent at any $C_{2v}$ symmetric configuration. As a result, the $^3\bar{B}_2$ structure undergoes a much weaker relaxation and forms a saddle point between the three equivalent $^3\bar{A}_2$ minima.} The equilateral triangle of N atoms is distorted only slightly, such that the N-N distances are 2.65 \AA\ , 2.65 \AA\ , and 2.67 \AA\ . The energy of the relaxed ${^{3}\bar{\text{B}}_2}$ configurations is 47~meV lower than the D$_{3\text{h}}$ symmetric ${^{3}\bar{\text{E}}^{\prime\prime}}$, which means that lowest energy barrier between three possible ${^{3}\bar{\text{A}}_2}$ symmetric Jahn-Teller minima is as high as $\delta_{\text{JT}} = $308~meV. Lastly, we note that none of the above-mentioned relaxed excited structures exhibit an out-of-plane component of the distortion, implying that the in-plane reflection symmetry of the structure remains unbroken.

The pseudo-Jahn-Teller effect and the configuration-dependent mixing of electronic states give rise to significant anharmonicity in the potential energy surface of the \VBm{} center. Consequently, instead of the slightly warped sombrero potential characteristic of the NV center, we obtain a tricorn hat-like potential, see Fig.~\ref{fig:fig2}b, in the $E \otimes e$ parameterization of the potential energy surface of \VBm. We note here that the parameters obtained from our calculations do not uniquely define an $E \otimes e$ potential energy surface, suggesting that this simple approximation does not perfectly describe this system. Taking a closer look at the structural relaxation induced by the pseudo-Jahn-Teller effect, we observe that the displacement of second-nearest-neighbor atoms contributes significantly to the relaxation, in contrast to the behavior of the NV center.

Since the 308~meV Jahn-Teller barrier energy of the \VBm{} center is significantly larger than the $k_{\text{B}}T$ at low temperatures, we conclude that the excited state of the  \VBm{} center is a static Jahn-Teller system for temperatures below $200$~K and exhibits a C$_{2v}$ symetric configuration, see Fig.~\ref{fig:fig2}b. However, we note that at elevated temperatures ($T > 200$~K), transitions between the JT minima occur within the lifetime of the triplet excited state, leading to temperature-dependent phenomena and transition to a dynamic Jahn-Teller system, as discussed later in the context of the excited state zero field splitting parameters. 

An important consequence of the JT distortion is that the optical transition with $\vec{E} \parallel z$ polarization axis becomes allowed~\cite{ivady_ab_2020,reimers_photoluminescence_2020}. We computed the radiative lifetime of the ${}^3\bar{\text{A}}_2$ state to be $\tau_{\text{rad}} = 98.8$~$\mu$s with WFT.  However, this value can be further reduced by symmetry-breaking fields~\cite{geng_deformation-driven_2025} and electron-phonon coupling~\cite{libbi_phonon-assisted_2022}. Indeed, Libbi~\emph{et al.} in Ref.~\citenum{libbi_phonon-assisted_2022} reported that not the static Jahn-Teller distortion, but the electron-phonon coupling with out-of-plane local vibrational modes is the most dominant factor in enabling the optical transition. Therefore, our value of $\tau_{\text{rad}} = 98.8$~$\mu$s is an upper limit. We note that the measurement of the radiative lifetime of the \VBm{} center is not straightforward, and the experiments are not conclusive so far~\cite{whitefield_magnetic_2023,baber_excited_2022,clua-provost_spin-dependent_2024}. For example, Clua-Provost~\emph{et al.}~\cite{clua-provost_spin-dependent_2024} attributed the decay of the optical excited state dominantly to non-radiative transitions.

In addition to the ${}^3\bar{E}^{\prime \prime}$ JT unstable state, there are an ${}^1\bar{E}^{\prime}$ and a ${}^1\bar{E}^{\prime \prime}$ JT systems in the singlet sector, further discussed in Supplemetary Note~3.


\subsection*{Zero-phonon line and phonon sideband}

According to previous theoretical studies~\cite{reimers_photoluminescence_2020} and our results, the dominant radiative relaxation pathway from the photoexcited \VBm{} center to the ground state occurs via the $^3\bar{A}_2 \rightarrow {}^3\bar{A}'_2$ optical transition. In the following, we discuss the spectroscopically relevant parameters of this process and compare them with available experimental data (see Table~\ref{table:PL_parameters} for a summary of the numerical results and Supplementary Note~1 for computational details).

The energy difference between the relaxed excited and ground states defines the zero-phonon line (ZPL), for which we obtain $E_{\text{ZPL}} = 1.88$~eV using our wavefunction-based methodology. \textcolor{black}{This value is slightly higher than our previous theoretical result of 1.62-1.71 eV~\cite{ivady_ab_2020} and also lies above the commonly observed energy range of \VBm{} photoemission. However, the latter is not necessary a contradiction.} The ZPL excitation energy is experimentally detectable only if it is not obscured by the phonon sideband (PSB). To assess the relative intensities of ZPL and PSB, we calculate the Huang-Rhys (HR) factor, $S$, defined as the ratio of the reorganization energy ($E_r = 0.29$~eV) to the energy of an effective phonon mode ($\hbar\omega_\text{eff} = 54$~meV) that mediates the geometry change between the ground and optically excited states. We obtain a Huang Rhys $S$ factor of 5.4, implying that, on average, nearly five phonons are emitted during radiative decay. This large HR factor originates from the substantial mass-weighted atomic displacement, $Q = 0.92~\sqrt{\text{amu}}\,\text{\AA}$, during the transition. The corresponding Debye-Waller (DW) factor,  $f_{\text{DW}}$, which quantifies the contribution of the ZPL peak to the total photoluminescence intensity, is as low as 0.4\%. \textcolor{black}{We note that in contrast to electronic energies, $\hbar\omega_\text{eff}$ and the resulting spectral lineshape converges slowly with model size, which made it necessary to investigate the largest affordable cluster model size (MODEL-5). As $f_{\text{DW}}$ shows a decreasing trend compared to MODEL-4, the aforementioned value of 0.4\% should be considered as an upper bound.}

These results are in good agreement with most experimental observations, according to which the photoluminescence spectrum of V$_\text{B}^-$ consists entirely of a broad PSB and the ZPL cannot be unequivocally identified~\cite{gottscholl_initialization_2020}. To provide a quantitative comparison with experiment, we examine the maximum of the phonon sideband, which appears at 1.46--1.60~eV in measurements~\cite{gottscholl_initialization_2020,reimers_photoluminescence_2020}. Theoretically, this maximum corresponds to the energy difference between the ground and excited states at the relaxed excited state geometry (i.e., $^3\bar{A}_2$). We calculate this energy to be 1.60~eV, reproducing the experimental PSB maximum within 0--0.14~eV, depending on the experimental data considered.

On the other hand, recent experiments on V$_\text{B}^-$ centers integrated into a high-$Q$ cavity reported enhanced emission intensities at 1.60~eV, which were subsequently assigned to the ZPL of the V$_\text{B}^-$ center~\cite{VB-ZPL}. We note that the theoretical energy difference of $E_r = 0.29$~eV between the calculated ZPL and the PSB maximum is approximately twice as large as the 0.14~eV separation between the ZPL and the PSB maximum reported in Ref.~\citenum{VB-ZPL}. The small Stokes shift of 0.14~eV would imply either (i) a low Huang-Rhys factor of 2.4 with a $\sim$59~meV phonon energy, and thus a larger DW factor of about 9\%, which would make the ZPL resolvable in other experiments as well, or (ii) phonon modes with much smaller energies, approximately 25.9~meV, together with a $S$ factor of 5.4, which fall outside the typical range of optical local vibrational modes. We therefore conclude that the peak observed in Ref.~\citenum{VB-ZPL} is probably not the ZPL of the V$_\text{B}^-$ center.

\begin{table}[h]
\setlength\extrarowheight{4pt}
\centering

\caption{\textcolor{black}{Characteristic parameters of the $^3\bar{E}'' \rightarrow {}^3\bar{A}'_2$ photoluminescence, obtained at CASSCF-NEVPT2 level.}}

\begin{threeparttable}
\begin{tabular}{|l|c|c|}
\hline
\multirow{2}{*}{Parameter (unit)} & \multicolumn{2}{|c|}{Theory}  \\
\cline{2-3}
& MODEL-4 & MODEL-5 \\
\hline
\hline
ZPL (eV) & 1.88\tnote{a} & -- \tnote{c}  \\
PSB$_\text{max}$  (eV) & 1.60\tnote{b} & -- \tnote{c}  \\
$Q$ ($\sqrt\text{amu}\mathrm{\AA}$) & 0.83 & 0.92 \\
$\hbar\omega_\text{eff}$ (meV) & 59 & 54 \\
$S$ (-) & 4.8 & 5.4\\
$f_{\mathrm{DW}}$  (-) & 0.008 & 0.004 \\
\hline
\end{tabular}
\begin{tablenotes}
\footnotesize
    \item[a] No reliable experimental value is known, see main text.
    \item[b] The experimentally observed range  is 1.46-1.60 eV. ~\cite{gottscholl_initialization_2020,reimers_photoluminescence_2020}
    \item[c] NEVPT2 energies were not obtained at MODEL-5 size due to unreasonably compotational costs. However, energies show signs of convergence at MODEL-4 size.
\end{tablenotes}
 
\end{threeparttable}

\label{table:PL_parameters} 
\end{table}

\section*{Singlet sector}

So far, we have discussed the properties of the triplet states, which define the magneto-optical properties of the V$_\text{B}^-$ center. The singlet excited states play a critical role in enabling spin initialization and spin readout, thus they also deserve a detailed description.

The lowest-energy $^{1}\bar{E}^{\prime}$ state in D$_{3\text{h}}$ symmetry is a mixture of two distinct configuration states. Approximately 78\% of the $^{1}\bar{E}^{\prime}$ state corresponds to the (1)$^{1}E^{\prime}$ configuration which is the zeroth-order excited singlet state of the ground state, meaning that the unpaired electrons are localized on the $e_{xy}^{\prime}$ state, similarly to the ground state. The remaining 22\% originates from the (2)$^{1}E^{\prime}$ pure group-theory state, which is a first-order excited configuration state where an electron is promoted from the $a_1^{\prime}$ state to the $e^{\prime}$ state. A similar effect, with a comparable degree of mixing, has been observed for the NV center in diamond, where it plays an important role in the lower-branch intersystem crossing decay.

The $^{1}\bar{E}^{\prime}$ state is also JT unstable and splits into a lower-lying $^{1}\bar{B}_1$ state \textcolor{black}{ and a $^{1}\bar{A}_1$ state} that is 92~meV higher in energy, which approximately defines the potential energy barrier between the JT minima. Since the barrier height is comparable to the 78~meV phonon energy in the singlet state, the $^{1}\bar{E}^{\prime}$ state is a dynamic JT configuration even at low temperatures

Next, we consider the second and third singlet excited states, \textcolor{black}{$^{1}\bar{\text{A}}_1^{\prime\prime}$} and $^{1}\bar{\text{E}}^{\prime\prime}$, respectively, which are found at approximately the same energy as their triplet counterparts. Similar singlet-triplet degeneracies, which are absent in bulk sp$^3$ host materials, have been reported in other hBN point defect systems\cite{babar_low-symmetry_2024}. The reason for this phenomenon lies in the coexistence of two symmetrically distinct sets of states, the $\sigma$ (single prime) states, which are symmetric with respect to in-plane reflection, and the $\pi$ (double prime) states, which are antisymmetric. Near-degeneracy is observed when the unpaired electrons in a triplet or open-shell singlet configuration occupy one $\sigma$ and one $\pi$ orbital. The resulting many-body state transforms according to a double-prime ($\pi$-like) irreducible representation, which is antisymmetric under in-plane reflection. Due to the strict orthogonality of the $\sigma$ and $\pi$ orbitals, the strong localization of the in-plane $\sigma$ states, and the partially delocalized nature of the $\pi$ states, the overlap of the $\sigma$ and $\pi$ states is minimal, which causes the exchange energy to be also minimal between the single and triplet many body states. However, it is important to note that the strict positivity of the exchange integral ensures that this difference is not zero, i.e.\ the singlet and triplet states are not degenerate by symmetry, but are accidentally close in energy, see Supplementary Note~4. This phenomenon has important consequences for the transition rates and leads to a decay mechanism different from that of the NV center.

Due to the similarity of the singlet manifold to the triplet optically excited state, the JT-unstable $^{1}\bar{\text{E}}^{\prime\prime}$ state undergoes a similar pseudo-JT distortion as the corresponding triplet excited state. The parameters of the singlet potential energy surface are equal to those of the triplet surface within the accuracy of our calculations, see Supplemetary Note~3. Consequently, the statement made about the temperature dependence of the $^{3}\bar{\text{E}}^{\prime\prime}$ state also applies to the $^{1}\bar{\text{E}}^{\prime\prime}$ state.

In contrast to the triplet sector, an optical emission with out-of-plane polarization is allowed for the ${}^1\bar{\text{E}}^{\prime\prime} \rightarrow {}^1\bar{\text{E}}^{\prime}$ transition. This selection rule also persists in the pseudo-Jahn-Teller distorted structure for the ${}^1\bar{\text{A}}_{2} \rightarrow {}^1\bar{\text{B}}_{1}$ transition, for which we calculate a ZPL energy of 1.41~eV using our wavefunction-based methodology. Despite the allowed transition, the calculated transition matrix element is small and consequently the radiative lifetime in the singlet optically excited state is long $\tau({}^1\text{A}_2) = 12$~$\mu$s.  
Similar to the triplet excited state, the Jahn-Teller and pseudo-Jahn-Teller effects induce significant structural deformations, resulting in a considerable distance, $dQ = 0.52\sqrt\text{amu}\mathrm{\AA}$, between the energy minima of ${}^1\bar{\text{A}}_{2}$ and ${}^1\bar{\text{B}}_{1}$. Consequently, we obtain a Huang-Rhys factor of 3.0 and a small Debye-Waller factor of 0.052. The maximum of the phonon-sideband emission is expected to occur in the near infrared region at 1.14~eV. In addition to the optical decay, we must consider non-radiative relaxation via internal conversion.
The ${}^1\bar{\text{A}}_{2} \rightarrow {}^1\bar{\text{B}_{1}}$ transition is symmetry-allowed and mediated by out-of-plane $b_2$ vibrational modes. Due to the reduced energy gap compared to the triplet transition, this relaxation pathway may effectively suppress the singlet optical transition.

\begin{figure*}[!h]
    \centering
    \includegraphics[width=\linewidth]{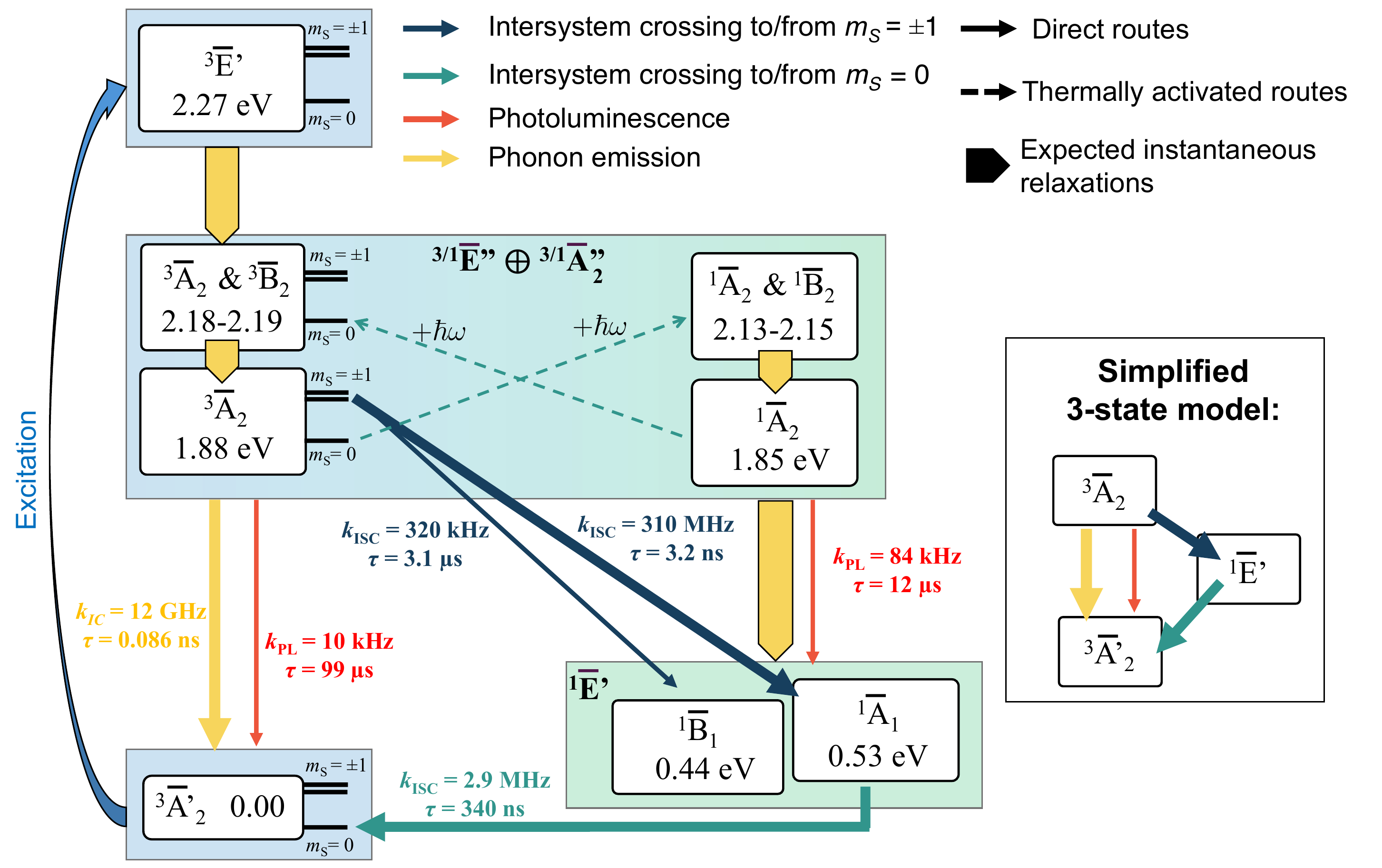}
    \caption{\textcolor{black}{Adiabatic transition energies and decay pathways of the \VBm{} center. The figure summarizes the transition rates ($k$) and the corresponding lifetimes ($\tau$)  between the lowest-lying electronic states, calculated using a 1D effective phonon approximation based on wavefunction theory data. The rates were obtained at the low-temperature limit (0 K) and at ambient pressure (0.1 MPa). For completeness, qualitatively characterized routes are also provided; pentagonal arrows, dashed arrows and unfilled arrows stand for suspected instantaneous nonradiative relaxations, thermally activated pathways and vibronically activated pathways, respectively.}}
    \label{fig:rates}
\end{figure*}

\section*{Transition rates between electronic states}

With the electronic structure and optimal geometries defined, we are now able to approximate the rates of the possible transitions that determine the ODMR signal of \VBm{} center using a 1D effective phonon model described in Supplementary Note 1. The results are summarized in Fig.~\ref{fig:rates}, while detailed numerical data, such as fitted potential energy surfaces and transition matrix element values along the 1D reaction coordinate, as the function of model size can be found in the Supplementary Tables. \textcolor{black}{We emphasize that even though the 1D effective phonon model has already been successfully applied for calculating PL, IC and ISC rates for color centers \cite{Smart2021,li_carbon_2022,PhysRevB.110.184302}, the maximal observed deviations of 1D-based rates from experiments occasionally approached an order of magnitude~\cite{Smart2021,PhysRevB.110.184302, Jin2025ISC}. Therefore, the presented results should be regarded as order-of-magnitude estimates, which are however suitable for} the qualitative mapping of interconversion rates among the low-lying many-body electronic states. A more rigorous quantitative investigation would require vibronic (rather than vibrational) states and multiple characteristic phonons, including out-of-plane vibrations. Such higher-level kinetic calculations will be addressed in future work.   

To understand how the ODMR signal arises, let us consider the possible transitions starting from the ${}^3\bar{A}_2$ state, which is the lowest energy triplet excited state over $^3\bar{A}'_2$.  The simplest possible relaxation is a direct conversion back to the ground triplet, $^3\bar{A}_2 \rightarrow {}^3\bar{A}'_2$. The rate of the radiative transition, photoluminescence, was found to be 10.1 kHz, corresponding to 98.8 $\mu s$ lifetime. 


Next, we turn our focus to intersystem crossing (ISC) towards the singlet spin sector. 
First, we examine the upper ISC branch that goes from the triplet optical excited state $^3\bar{\text{A}}_2$ to the singlet sector. There are two possible directions: decay to the lowest-lying JT split ${}^1\bar{\text{E}^{\prime}}$ or to the JT split ${}^1\bar{\text{E}^{\prime\prime}}$ allowed dominantly with $\lambda_{\perp} = 450$~GHz matrix element and $\lambda_{\parallel} = 4.7$~GHz matrix element, respectively. This means that the $m_{\text{S}} = \pm 1$ and the $m_{\text{S}} = 0$ states can decay via two distinct pathways in stark contrast to the NV center in diamond. 

Despite the significant energy gap between the  $^3\bar{\text{A}}_2$  and the JT split ${}^1\bar{\text{E}^{\prime}}$, we found that the $m_{\text{S}} = \pm 1$  decay channel is fast. Using the 1D effective phonon model, we obtain $\tau({}^3\text{A}_2, m_{\text{S}}) = 3.2$~ns, which is of the same order of magnitude as in the experiments\cite{whitefield_magnetic_2023,clua-provost_spin-dependent_2024}. We note that the $\lambda_{\parallel}$ component, which would enable direct decay to the split ${}^1\bar{\text{E}^{\prime}}$ state, is negligibly small in our calculations, see more details in Supplementary Table~10.

Regarding the lower ISC branch $^1\bar{E}' \rightarrow {}^3\bar{A}_2$, which closes the spin polarization cycle, we found that ISC is symmetry-allowed only starting from one of the $C_{2v}$-symmetric Jahn-Teller states, $^1\bar{A}_1$. This state is, however, populated by the previously discussed dynamic Jahn-Teller effect, which constantly interconverts the $^1\bar{A}_1$ and $^1\bar{B}_1$ states. The 1D effective phonon model predicts a \textcolor{black}{2.9~MHz} rate for the decay to the $^3\bar{A}_2$ ground state, and therefore the $^1\bar{E}'$ state acts as a metastable shelving state with a lifetime of approximately 340~ns. Recent measurements reported an order-of-magnitude shorter lifetime for the longest-lived metastable singlet~\cite{whitefield_magnetic_2023,clua-provost_spin-dependent_2024}. This discrepancy may have originated from temperature effects ignored in our model but potentially relevant for the $^1\bar{\text{E}}$ state, or from the low-level approximation inherent to the 1D effective phonon model. Owing to the strict selectivity of the lower-branch ISC, we expect close to complete polarization in the $m_S = 0$ sublevel upon optical pumping.

\textcolor{black}{The hitherto discussed routes provide no explanation for the relaxation to the ground state from the $m_S = 0$ sublevel of $^3\bar{A}_2$, even though the absence of such a pathway clearly contradicts the experimentally measured nanosecond-scale overall lifetime of the triplet excited state. One possibility is a vibronically allowed ISC outside $C_{2v}$ symmetry in the direction of $^3\bar{A}_2 \rightarrow ^1\bar{E}'$ (Fig. 3, unfilled arrows); nevertheless, as the corresponding matrix elements are zero in both the initial-state and final-state equilibrium geometries, it is highly unlikely that such a vibronic process can proceed at comparable rate to the strongly allowed $m_S = \pm 1$ channel. The two remaining alternatives are (1) a scattering between quasi-degenerate triplets and singlets, (2) direct nonradiative IC to $^3\bar{A}'_2$ via out-of-plane phonons. These routes are discussed in detail in the following.  } 

Concerning  the singlet-triplet quasi-degeneracy, the transition between the pseudo JT relaxed ${}^3\bar{\text{A}}_2$  and the ${}^1\bar{\text{A}}_2$ state spin orbit matrix element is not obtained for symmetry reasons. On the other hand, the ${}^3\bar{\text{A}}_2 \leftrightarrow {}^1\bar{\text{B}}_2$ as well as \textcolor{black}{the ${}^3\bar{\text{B}}_2 \leftrightarrow  {}^1\bar{\text{A}}_2$ transitions} are allowed by SOC matrix element of 4.7~GHz. Since the $\bar{\text{B}}_2$ states are higher in energy, these transitions are enabled by quantum effects, such as zero-point fluctuation and thermally excited phonons (see dashed lined in Fig. 3.). Modeling this transition requires the development of a multi-surface multi-phonon dynamic simulation method beyond the Born-Oppenheimer method, which is subject to subsequent work. Despite these uncertainties due to the lack of a proper method, we can make simple qualitative statements. The large coupling matrix element and the proximity of the states in energy, one could reasonably expect a fast triplet-singlet transition. Owing to the expected involvement of thermally excited phonons and the static to dynamic JT transition with temperature, we anticipate that the $m_{\text{S}} = 0$ transition rate is highly temperature and potentially strain-dependent. When transitioned to the single pseudo JT ${}^1\bar{\text{E}}^{\prime\prime}$ system, the system is expected to decay to the JT ${}^1\bar{\text{E}}^{\prime}$ state via fast internal conversion (which is forbidden in 1 dimension under $C_{2v}$ symmetry constraints, but is expectedly allowed by out-of-plane phonons). Nevertheless, owing to the quasi degenerate singlet-triplet excited electronic structure, it is also possible that after a transition to the pseudo JT ${}^1\bar{\text{E}}^{\prime\prime}$ state, the system transforms back to the triplet sector. Although this backscattering can be suppressed by a fast decay to the ${}^1\bar{\text{E}}^{\prime}$ state, it could, in principle, lead to elongated PL timescales. 

The $^3\bar{A}_2 \rightarrow {}^3\bar{A}'_2$ relaxation via internal conversion seems counterintuitive at first sight, as the energy gap of 1.88 eV is large and the transition matrix elements are zero in the 1D effective phonon model (due to the different irreducible representation of the initial and final states at $C_{2v}$ symmetry). Nevertheless, to estimate the order of magnitude of this rate, we ran multiphonon excited-state dynamics (ESD) calculations for this sole transition, which also accounts for out-of-plane vibrations. The calculation was performed with the ESC machinery of the ORCA software at time-dependent density functional theory (TD-DFT) level, as WFT-level nonadiabatic couplings are not yet available. We confirme, however, that the triplet TD-DFT and CASSCF  wavefunctions are closely related (see SI, Supplementary Note 5). Interestingly, we obtained a rate of 12 GHz (i.e., 0.086 ns lifetime), which even exceeds that of the competing ISC pathway. As IC is not affected by the $m_S$ quantum number, it is reasonable to assume that this $^3\bar{A}_2 \rightarrow {}^3\bar{A}'_2$ conversion is the primary pathway of relaxation at the $m_S = 0$ channel. 

\textcolor{black}{Summarizing the results above, we can conclude that the magneto-optical behavior of the \VBm{} center can be well-characterized by a three-state model, as highlighted in Fig.~3, right. Upon optical excitation, the $^3\bar{A}_2$ state starts to accumulate, the weak photoluminescence of which ($^3\bar{A}_2 \rightarrow ^3\bar{A}'_2$) can be observed by optical detectors. The quantum yield of the defect is severely reduced by sub-nanosecond-range internal conversion to the order of $10^{-6}$. An alternative relaxation route involving the lowest singlet electronic state (i.e., the dynamic Jahn-Teller state of $^1\bar{E}'$) competitively consumes $^3\bar{A}_2$ at the $m_S = \pm 1$ chanel and produces the triplet ground state with $m_S = \pm 0$ sublevel, that is, the ISC-mediated pathway practically flips $m_S$. This three-state model implies nanosecond-scale excited-state lifetime, ground-state polarizability towards $m_S = 0$, as well as the emergence of ODMR contrast due to the high selectivity of both ISC branches, in complete agreement with experiments.     }

\textcolor{black}{
Finally, we compare our results with the pioneering study of Ref.~\citeonline{reimers_photoluminescence_2020}, which described the \VBm{} spectrum using a combination of electronic-structure approaches, complemented by a posteriori corrections for environmental and basis-set effects, and semiclassical approaches to estimate interconversion rates.
For the triplet manifold, Reimers et al. used equation-of-motion coupled-cluster theory with singles and doubles (EOM-CCSD) at CCSD/CAM-B3LYP-optimized geometries, motivated by the  single-reference character of the triplet ground state.
The strongly correlated open-shell singlet states were treated using double-ionization-potential EOM-CCSD (DIP-EOM-CCSD), with their geometries optimized at the CASSCF level.
For a direct comparison, we consider the vertical excitation energies of MODEL-2, which is the largest model studied  by the CC variants. The reported EOM-CCSD relative energies reproduce our CASSCF-NEVPT2 results to within 0.1 eV for the triplet states. The DIP-EOM-CCSD results show differences of about 0.2 eV on average for the singlet states. Similarly,  the final   predictions of Ref.~\citeonline{reimers_photoluminescence_2020} on the low-energy adiabatic spectrum is consistent with our findings obtained for MODEL-4 , with differences of approximately 0.1–0.3 eV.
Furthermore, we find that the lifetimes of the optical excitations and the dominant upper branch of the ISC are also well comparable, agreeing within a factor of  1.2 to 9.
In contrast,  the study reported an extremely long  singlet recovery time ($10^{23}$ s) with substantial uncertainty arising from the underlying modeling approximations.
According to our findings based on the  effective-phonon theory, the $^1\bar{\text{A}}_1$ state relaxes to the ground state with a lifetime of 340~ns, consistent with experimental findings.
Interestingly, Reimers et al. also predicted the IC from $^3\bar{A}'_2$ to the ground state to be practically forbidden according to transition-state theory, whereas our explicit multiphonon excited-state dynamics simulations yield a finite lifetime comparable in magnitude to those of the ISC processes.
}

\begin{table}[h]
\setlength\extrarowheight{4pt}
\centering

\caption{Zero-field splitting parameters of triplet electronic states. Theoretical values are obtained on wavefunction-based theory at MODEL-4 size (see the Methods), while the corresponding experimental values are from the literature. 
}

\begin{threeparttable}
\begin{tabular}{|c|cc|cc|}
\hline
State  & $D_\mathbf{calc}$ (GHz) & $D_\mathbf{exp}$ (GHz) & $E_\mathbf{calc}$ (MHz) & $E_\mathbf{exp}$ (MHz) \\
\hline
\hline
$^3\bar{A}'_2$ & +3.38 & 3.63\tnote{a} & 0.00 & 0--150\tnote{b}   \\
$^3\bar{A}_2$ & +1.91 & 2.11\tnote{c} &  465 & \textcolor{black}{$258\pm 2$}\tnote{c} \\ 
\textcolor{black}{$^3\bar{A}''_1$} & +2.88 & -  & 0.00 & - \\
\hline
\end{tabular}
\begin{tablenotes}
\footnotesize
    \item[a] Ref.~\citenum{gottscholl_spin_2021}  
    \item[b] Refs.~\citenum{Durand-2023,Carbone-2025}; the reported values cover a wide range.
    \item[c] Ref.~\citenum{Mathur2022}, based on the ODMR spectrum at 10 K, only the absolute value is known.
\end{tablenotes}
\end{threeparttable}

\label{table:ZFS_data}  
\end{table}

\section*{\textcolor{black}{Simulation of spin polarization and ODMR contrast}}

\textcolor{black}{Having obtained the IC, ISC and PL rates of the three-state model highlighted in Fig.~\ref{fig:rates}, we can simulate the spin polarization and the ODMR cycle of the \VBm{} center and compare the results directly with experiments. To this end, we consider the three spin sublevels of the $^3\bar{A}'_2$ ground state and of the $^3\bar{A}_2$ excited state and the $^1\bar{E}'$ shelving state, and describe their dynamics by a Lindblad master equation for the density matrix. The resonant microwave (MW) drive of both ground-state spin transitions with Rabi frequency $\Omega$ is included coherently, while the spin-conserving optical excitation with rate $k_{\text{abs}}$, the calculated PL, IC and ISC processes, the ground-state spin-lattice relaxation with $T_1 = 18$~$\mu$s~\cite{gottscholl_2020}, and the ground-state spin dephasing characterized by $T_2^*$ enter as incoherent processes. The ODMR contrast is obtained from the steady state of the master equation, while in the absence of MW drive the master equation reduces to classical rate equations, which provide the degree and timescale of the spin polarization. 
We note that, owing to the 1D effective phonon approximation, the calculated rates and the derived quantities should be considered as order-of-magnitude estimates.}

\textcolor{black}{As the calculated IC rate dominates the decay of the optically excited state and it is the most uncertain quantity of our model, we performed the simulations with two parameter sets. In the first set, all rates are taken from our \emph{ab initio} calculations. In the second set, the IC rate is set to $k_{\text{IC}} = 8.3\times10^{8}$~s$^{-1}$, which reproduces the measured $\approx1.2$~ns lifetime of the $m_S=0$ excited sublevel\cite{clua-provost_spin-dependent_2024,Lee2025}, while all other rates are kept at their calculated values. In both cases, the MW drive is characterized by $\Omega/2\pi = 10$~MHz and $T_2^* = 13.8$~ns\cite{Nakamura2026}, and $k_{\text{abs}} = 10^7$~s$^{-1}$ is used as a representative pump rate of moderate confocal excitation. The results are summarized in Table~\ref{tab:odmr_comparison}.}

\textcolor{black}{Both parameter sets reproduce the spin selectivity of the optical cycle, i.e. optical pumping into the $m_S=0$ ground-state sublevel, a shorter excited-state lifetime for $m_S=\pm1$ than for $m_S=0$, and a decrease of the photoluminescence under resonant MW drive. The magnitude of these effects, however, depends strongly on the IC rate. With the \emph{ab initio} IC rate, IC dominates the decay of all excited sublevels, therefore only $\eta = 2.6\%$ of the $m_S=\pm1$ excitations decay through the ISC channel (ISC branching ratio), and the lifetimes of the $m_S=0$ and $m_S=\pm1$ excited sublevels differ by only 3\%. Consequently, the low-power limit of the ODMR contrast, which is determined by the ratio of these lifetimes, is as small as $C_0 = 1.8\%$, and the steady-state spin polarization reaches only 88\% as the optical pumping competes with the ground-state spin-lattice relaxation. When the IC rate is set to reproduce the measured $m_S=0$ lifetime, the ISC branching ratio increases to 27\%, the steady-state polarization of 98.7\% becomes consistent with the experimentally inferred $\geq95\%$~\cite{Lee2025}, and the CW-ODMR contrast increases to 47\%. The contrast is enhanced by the accumulation of population in the long-lived $^1\bar{E}'$ state, and it decreases to 23\% if the measured 44~ns singlet lifetime\cite{Lee2025} is used instead of the calculated 340~ns. For comparison, the highest reported CW-ODMR contrast of 46\% was achieved with plasmonic enhancement\cite{gao_high-contrast_2021}. The polarization time scales inversely with $\eta k_{\text{abs}}$. The estimated $t_{95} = 109$~ns~\cite{Konrad2026} polarization time is reproduced at $k_{\text{abs}} \approx 1.4\times10^{8}$~s$^{-1}$ in the second parameter set. The remaining deviations, i.e. the longer $m_S=\pm1$ excited-state lifetime (0.87~ns versus $\approx0.5$~ns) and the longer singlet lifetime (340~ns versus 44~ns), indicate that our 1D effective phonon model underestimates both ISC rates by a factor of 4 to 8. Overall, the comparison suggests that the calculated IC rate is overestimated by about one order of magnitude, while the three-state model with a corrected IC rate consistently explains the degree and timescale of the spin polarization and the emergence of a large ODMR contrast.}

\begin{table}[h]
\centering
\small
\setlength{\tabcolsep}{3pt}
\caption{\textcolor{black}{Calculated and experimental kinetic properties of the polarization and ODMR cycle of \VBm{} center. In parameter Set 1, all rates are obtained from \emph{ab initio} calculations, while in parameter Set 2, the IC rate is adjusted to reproduce the measured lifetime of the $m_S=0$ excited sublevel. The ODMR contrasts are obtained from the steady state of the Lindblad master equation. The polarization and ODMR quantities are calculated for $k_{\text{abs}}=10^7$~s$^{-1}$, $T_1=18$~$\mu$s~\cite{gottscholl_2020}, $T_2^*=13.8$~ns~\cite{Nakamura2026}, and $\Omega/2\pi=10$~MHz drive of ground-state spin transitions. $C_0$ is the low-power limit of the contrast with saturated MW drive.}}
\label{tab:odmr_comparison}
{
\begin{tabular}{|l|c|c|c|}
\hline
Property & \makecell{Set 1\\(\emph{ab initio} IC)} & \makecell{Set 2\\(exp. IC)} & \makecell{Experiment /\\estimate} \\
\hline
\multicolumn{4}{l}{\textit{Process-specific lifetimes (ns)}} \\
\hline
PL, $^3\bar{A}_2\to {}^3\bar{A}'_2$
  & $99000$ & $99000$ & --- \\

IC, $^3\bar{A}_2\to {}^3\bar{A}'_2$
  & $0.086$ & $1.2$ & --- \\

ISC, $^3\bar{A}_2(m_S=\pm1)\to {}^1\bar{E}'$
  & $3.2$ & $3.2$
  & $0.778$~\cite{Lee2025} (fit) \\

ISC, $^1\bar{E}'\to {}^3\bar{A}'_2(m_S=0)$
  & $340$ & $340$
  & $44.1$~\cite{Lee2025} (fit) \\

Excited-state lifetime, $m_S=0$
  & $0.086$ & $1.2$
  & \shortstack{
      $1.22\pm0.09$~\cite{clua-provost_spin-dependent_2024}\\
      $1.18\pm0.09$~\cite{Lee2025}} \\

Excited-state lifetime, $m_S=\pm1$
  & $0.084$ & $0.87$
  & \shortstack{
      $0.53\pm0.06$~\cite{clua-provost_spin-dependent_2024}\\
      $0.47\pm0.05$~\cite{Lee2025}} \\

ISC branching ratio, $\eta$
  & $2.6\%$ & $27\%$ & $\approx60\%$~\cite{Lee2025} (derived) \\

\hline
\multicolumn{4}{l}{\textit{Polarization degree and timescale}} \\
\hline
Ground-state $m_S=0$ polarization
  & $88.4\%$ & $98.7\%$
  & $\geq95\%$~\cite{Lee2025} (inferred) \\

$t_{95}$ ($\mu$s)
  & $9.4$ & $1.1$
  & $0.109$~\cite{Konrad2026} (model) \\

$k_{\text{abs}}$ for $t_{95}=109$~ns (s$^{-1}$)
  & $1.5\times10^{9}$ & $1.4\times10^{8}$ & --- \\

\hline
\multicolumn{4}{l}{\textit{ODMR contrast}} \\
\hline
Low-power limit, $C_0$
  & $1.8\%$ & $18\%$ & --- \\

Continuous-wave ODMR contrast
  & $6.0\%$ & $47\%$
  & $46\%$~\cite{gao_high-contrast_2021} (plasmonic) \\
\hline
\end{tabular}
}
\end{table}

\section*{Zero-field splitting parameters of triplets}
Next, we compare the experimental $D$ and $E$ tensors~\cite{Mathur2022}, gathered in Table~\ref{table:ZFS_data}, to the theoretical values, obtained from the relative energy levels of spin sublevel dependent eigenstates (see more details on the introduction of spin-orbit coupling to WFT calculations in Supplementary Note 1).

We begin the discussion with the $^3\bar{A}'_2$ ground state. As it has already been demonstrated, this is a single-reference system, for which DFT methods would also be appropriate. However, the computation of its zero-field splitting parameters by CASSCF-NEVPT2 provides information on the error margin of the method. For the $D$ splitting, we obtain 3.38 GHz, which is close to the experimental low temperature value of $D_{\text{GS}}^{\text{exp}} = 3.63$~GHz\cite{gottscholl_spin_2021}. The $E$ tensor is zero in high symmetry, as the $m_S = \pm 1$ states are split neither by spin-orbit nor by spin-spin coupling. This finding also corresponds well to the low reported $E$ tensors of 0-150 MHz~\cite{Durand-2023,Carbone-2025}. The observation of a $E$ finite splitting demonstrates the presence of an external electric field, caused by e.g.\ other defects in the lattice~\cite{Udvarhelyi-2023}.      

Having verified that zero-field splitting (ZFS) parameter calculations are reliable, we continue with the characterization of excited states.  The pseudo Jahn-Teller effect of the optical excited states has a considerable influence on the excited-state zero-field splitting tensor. Owing to the C$_{2v}$ symmetry, one would expect a significant transverse zero-field splitting, quantified by the $E_{\text{ES}}$ parameter, which splits the $m_{\text{S}} = \pm 1$ states in the excited state. For the pseudo Jahn-Teller distorted $^{3}\bar{\text{A}}_{2}$ states we obtain $D_{\text{A}_2}= 1.91$~GHz \textcolor{black}{and $E_{\text{A}_2}= 465$~MHz} ZFS splitting parameters on QDPT CASSCF-NEVPT2 level of theory at zero temperature in the classical sense, i.e.\ without including zero point energy of the local vibrational modes.  We find the quantization axis to be parallel to the $c$ axis of the hBN lattice in the static JT configuration. The experimental excited state ZFS parameter of \textcolor{black}{D$_{\text{ES,exp}} = 2.11 \pm 0.03$~GHz} \cite{Mathur2022} is in good agreement with the theoretical value of $D_{\text{A}_2} = 1.91$~GHz.

For completeness, we also computed the $D$ tensor of \textcolor{black}{$^3\bar{A}''_1$}, which we found to be 2.88 GHz. Conspicuously, neither this $D$ value, nor the symmetry-forbidden splitting of $\pm 1$ sublevels, i.e.\ $E = 0$ MHz for \textcolor{black}{$^3\bar{A}''_1$}, match the experimental observation, which further confirms ordering of the states in Fig.~\ref{fig:fig2} and the identification of the JT split $^3\bar{E}''$ state as the source of the excited state ODMR signal.

Next, we study the temperature dependence of the excited-state ZFS parameters by calculating the occupation dynamics of the different JT minima, assuming temperature-dependent transitions between them. Since the principal axis of the ZFS tensor is perpendicular to the plane in all minima, the component $D$ with corresponding eigenvector parallel to the $c$ axis is not affected by transitions between the JT minima. Thus, to first order, $D(T) \sim \text{const}$, in agreement with experimental observations.\cite{Mathur2022} Due to anharmonic effects and lattice expansion, smaller deviations are observed.\cite{Mathur2022} The transverse zero-field splitting parameter, however, is strongly affected by the JT dynamics when transitions occur, as the in-plane eigenvectors of the ZFS tensor corresponding to the transverse splitting point in different directions in each minimum. Frequent transitions between the minima can completely average out the transverse component and, as a result, yield a high-symmetry-like fine structure in the excited state.

To study this process, we approximate the tunneling rate using an Arrhenius formula (see Supplementary Note~3) and estimate the time-averaged transverse zero-field splitting parameter. The temperature dependence of the transverse zero-field splitting parameter is shown in \textcolor{black}{Fig.~\ref{fig:fig2}c } together with available experimental data.\cite{Mathur2022} As can be seen, both the experimental and theoretical results exhibit a strong temperature dependence. Based on the simulations, we identify three temperature regimes with distinct dynamics. Below $\sim$200~K, the excited state behaves as a static JT-distorted configuration. Between $\sim$200~K and $\sim$400~K, transitions between the JT minima occur within the lifetime of the excited state, taken to be 1.7~ns in our analysis, with increasing rates that partially average out the transverse ZFS parameter. Above $\sim$400~K, the transitions become so frequent that the JT-induced $E$ parameter is completely averaged out. We note that in the transition region between 200~K and 400~K, we expect an increase of the spin noise and a shortening of the spin lifetime, which could give rise to the spin resonance line broadening observed in Ref.~\citenum{Mathur2022}.

Mathur \textit{et al.}\,\cite{Mathur2022} measured the excited-state transverse ZFS parameter to be \textcolor{black}{$E_{10\,\text{K}} = 258 \pm 2$~MHz} at 10~K and \textcolor{black}{$E_{\text{RT}} = 74\pm 42$~MHz} at room temperature. Our calculated values, $E_{<200\,\text{K}} = 465$~MHz and $E_{298\,\text{K}} = 112$~MHz, are approximately 80\% and 50\% larger than the corresponding experimental values, respectively.

\textcolor{black}{We note that the absolute deviation of the calculated $E$ (207 MHz) is comparable to the deviation of the $D$ parameters of both the ground and excited states. The large relative error mainly originates from the small magnitude of $E$. Unlike $D$, however, the transverse ZFS parameter transforms as the $e'$ Jahn-Teller distortion and depends linearly on its amplitude, which is reflected in its larger cluster-size and strain sensitivity (Supplementary Table 11). Consequently, a moderate overestimation of the pseudo Jahn-Teller distortion by the CASSCF geometry optimization, together with the neglected vibrational averaging along the $e'$ mode, directly translates into an overestimated $E$, while $D$ remains largely unaffected.} When employing experimental low-temperature values in our Arrhenius analysis, however, we achieve good agreement with the experimental high-temperature value of the transverse zero-field splitting parameter.


%
\begin{table*}[!th]
\setlength\extrarowheight{4pt}
\centering

\caption{\textcolor{black}{Horizontal and vertical compression dependence of zero-field splitting parameters and characteristic transition rates.  The provided derivatives were determined based on a linear fit to 5 data points (see SI, Supplementary Note 6 for raw numerical data). The calculations were performed at 3-layer MODEL-2 size for specific $p$ compression values as discussed in the main text. 
 }
}

\begin{threeparttable}

\begin{tabular}{|>{\centering}m{4.5cm}|l|c|c|c||c|c|c|}
\hline
\multirow{2}{4.5cm}{\centering Property type (unit)}&\multirow{2}{*}{Parameter} & \multicolumn{2}{c|}{Sensitivity to compression \tnote{a}} \\
\cline{3-4}
& &  Horizontal & Vertical  \\
\hline
 \hline
\multirow{4}{4.5cm}{\centering Zero-field splitting (MHz/GPa)} & $D\,(^3\bar{A}'_2)$ &  +38\tnote{b}  & -4 \\
&$E\, (^3\bar{A}'_2)$ & 0 & 0   \\
& $D\, (^3\bar{A}_2)$ &  +11  & -11  \\
& $E\, (^3\bar{A}_2)$  &  +23  & +11  \\
 \hline
\multirow{2}{4.5cm}{\centering Relative transition rates (\%/GPa)}& 
PL ($^3\bar{A}_2 \rightarrow ^3\bar{A}'_2$) &   +2  & 0   \\
& ISC ($^3\bar{A}_2 \rightarrow ^1\bar{A}_1$) &  +5   \tnote{d} & +19   \tnote{d} \\
\hline

\end{tabular}

\begin{tablenotes}
    \footnotesize
    \item[a] First-order derivative calculated at ambient (0 GPa) pressure. 
    \item[b] From Ref.~\citenum{mu2025magneticimaginghighpressure} the experimental value is 48$\pm$1 MHz/GPa
    \item[c] Transition matrix element averaged between initial-state and final-state geometries.
    \item[d] Experimental study~\cite{mu2025magneticimaginghighpressure} reports the increase of \textcolor{black}{nonradiative rates} under pressure without quantitative details.
\end{tablenotes}
\end{threeparttable}

\label{table:Pressure_data}  
\end{table*}

\section*{Strain dependence of \textcolor{black}{magneto-optical properties}}

Based on the kinetic model established in the previous sections, it becomes possible to theoretically interpret the recently reported strain-sensing capabilities of the \VBm{} center in hBN~\cite{mu2025magneticimaginghighpressure}. Experimentally, the application of a hydrostatic pressure of a few GPa was observed to cause (i) a slight positive shift of the D tensor and (ii) a decrease in photoluminescence intensity~\cite{mu2025magneticimaginghighpressure}. Upon unidirectional strain, a decrease of the ZFS $D$ value and an increase of the $E$ value were observed~\cite{Udvarhelyi-2023}. In this section, we model the effect of strain using deformed molecular models, that is, the positions of the outer, fixed atoms were adjusted according to the applied strain, and we repeated the CASSCF-NEVPT2 electronic structure calculations, including relaxation of the inner atoms under this constraint. \textcolor{black}{This approach is expected to reproduce the experimentally observable pressure dependence qualitatively, even though quantitative agreement may require larger models to capture the long-range elastic response of the surrounding lattice.}

The effect of isotropic pressure manifests itself in two main structural changes. First, the distance between individual hBN layers decreases, as dictated by the elastic constant $C_{zz} = 27$~GPa~\cite{Bosak-2006}. Accordingly, an external pressure of 2.7~GPa is required to shrink the interlayer distance (3.33~{\AA} at ambient pressure) by 10\%. Naturally, this effect, which will be referred to as ``vertical compression'' in the following, can only be studied using three-layer models. The size of MODEL-2 was used for these calculations to keep the computations feasible. To quantitatively investigate the pressure dependence, we set five different interlayer distances at 90\%, 95\%, 100\%, 105\%, and 110\% of the original value, corresponding to external pressures of 2.7~GPa, 1.35~GPa, 0~GPa, $-1.35$~GPa, and $-2.7$~GPa, respectively. While negative pressures are not physically meaningful, they help to obtain a better linear fit.
 
Secondly, the individual layers themselves shrink. This deformation, which will be referred to as "horizontal compression", is, however, less profound in geometry. Namely, the in-plane elastic tensor elements of $C_{xx}=C_{yy}=811$~GPa is 30 times larger than $C_{zz}$~\cite{Bosak-2006}. Thus, to mimic the effect of the aforementioned pressure range, the distance of outer (fixed) atoms to the vacancy was set to 99.67\%, 99.83\%, 100\%, 100.17\%, and 100.33\% of the original value, respectively, before recomputing all properties at MODEL-2 size. The results of our investigations for all data points are detailed in Supplementary Note 6 of the SI, while Table~\ref{table:Pressure_data} summarizes the observed main trends. In our study, the effects of in-plane (horizontal) and out-of-plane (vertical) strain can be compared, which helps identify the leading terms in the pressure and strain dependence.

In the top rows of Table~\ref{table:Pressure_data}, we address the change of zero-field splitting parameters. As for the ground-state D tensor, the in-plane distorted model shows an unequivocally increasing splitting under pressure, compared to which the out-of-plane strain component has a negligible effect. Therefore, even though the vertical strain is of the order of magnitude larger at isostatic pressure, the change of the ground state D tensor is governed by the smaller in-plane distortions. This is presumably due to the strongly localized in-plane dangling bonds forming the unpaired defect states. This conclusion is in line with previous theoretical results of single-axis strain studies by Udvarhelyi et.al.\cite{Udvarhelyi-2023}. Note that the calculated 38~MHz/GPa slope of the linear fit of pressure dependence qualitatively agrees with experimental 48 $\pm$ 1 MHz/GPa~\cite{mu2025magneticimaginghighpressure}. The remaining discrepancy presumably results from the limited model size. 

Interestingly, when investigating the effect of pressure on the $D$ and $E$ parameters of the ${}^3\bar{\text{A}}_2$ excited state, we observed reduced sensitivity to the inplain stress, for example, the sensitivity of the D tensor of $^3\bar{A}_2$ on $p_{xy}$ is only +11 MHz/GPa. Furthermore, as the derivative with respect to perpendicular stress $p_z$ has commensurable value but opposite sign. Consequently, the optical excited-state ZFS parameters are less strain sensitive.

\textcolor{black}{Next, we examined the stress dependence of two key transition rates which determine the measurable photoluminescence intensity. The radiative decay (PL) can be considered a practically pressure-independent process: it is completely insensitive to $p_z$ and its rate is expected to increase by only about 2\% per GPa of applied $p_{xy}$ pressure. In sharp contrast to this, ISC for the ${}^3\bar{\text{A}}_2 \rightarrow {}^1\bar{\text{A}}_1$ decay channel allowed for the $m_S = \pm1$ state turned out to be significantly accelerated by increasing pressure (+5\%/GPa and +19\%/GPa for horizontal and vertical compression, respectively). This can be partly traced back to the compression-enhanced SOC matrix elements, and partly to enhanced phonon overlap via stiffer phonons and the decrease of the energy gap.}

Our numerical results properly reflect the decrease in the excited state lifetime and PL intensity under hydrostatic pressures, observed in experiments~\cite{mu2025magneticimaginghighpressure}. Namely, the enhanced $m_S = \pm1$ ISC rate and the nearly unchanged radiative decay imply both the decrease of lifetime and the quantum yield of photoluminescence. Altogether, we can attribute the observed pressure dependence of the ground-state D tensor and the excited-state dynamics to the horizontal and vertical compression of the hBN lattice, respectively.

\bigbreak
\section*{\large Discussion}
\label{sec:discussion}

In summary, we have presented a comprehensive multi-reference study of the \VBm{} center using the CASSCF-NEVPT2 computational approach. By accurately considering both correlation and relaxation effects, we systematically investigated the low-lying excitations, transition energies, zero-field splitting parameters, and the rates of photoluminescence and intersystem crossing. Our theoretical predictions demonstrate good agreement with available experimental findings for ZFS and photoluminescence parameters.

Despite these, our findings underscore that the theory of the \VBm{} optical cycle is not complete yet. The quantitative description of the nonradiative decay pathways, specifically the transition rates from the $m_S = 0$ channel to the singlet sector, exhibits significant difficulties due to the pseudo-Jahn-Teller quasi-degenerate singlet-triplet states involved. Because of these, the transitions are deeply influenced by electron-phonon interaction and quantum effects. The \textcolor{black}{$E \otimes e$} JT formalism and the 1D effective phonon model developed for the NV center are insufficient here, and development of a multi-surface dynamic simulation method beyond the Born-Oppenheimer approximation is needed to fully unravel excited state dynamics.

Due to the complexity of the ${}^3\bar{\text{E}} \oplus^1\bar{\text{E}}$ states, it is expected that the transition rate depends on all factors that influence the potential energy surface, population of the vibronic state, and the splitting of the singlet-triplet quasi-degenerate structure. Consequently, temperature and strain dependence can play a crucial role here, enabling, in turn, an engineering of the relaxation mechanism and the ODMR contrast.

Furthermore, our exploration of the defect's higher-lying excited states points towards similar avenues by enabling multiple ways of excitation, both in the triple and singlet sectors. Such dynamics may be probed by two laser-pulse schemes potentially leading to reverse decay from the singlet excited states to the triplet excited state upon pumping the singlet ${}^1\bar{B}_1 \rightarrow {}^1\bar{A}_2$, which could enhance both luminosity and contrast. 

Further understanding these intricacies is essential for the advancing quantum metrology with the \VBm{} center. 

\section*{\large Methods}
\label{sec:tech}

\subsection*{Ab initio methods}

In this work, we focus on the discussion of the photoluminescence properties obtained for the \ce{B60N60H27-} cluster model of the \VBm{} color center depicted in Supplementary Figure~1. Excitation energies and relaxation effects were obtained using the complete active space self-consistent field (CASSCF)~\cite{Roos1987,Olsen-2011,casscf} approach improved by $n$-electron valence perturbation theory (NEVPT2)~\cite{Angeli-2001a,Angeli-2001,Angeli2007,Guo_2021,Kollmar_2021} as summarized below with example input files in Supplementary Note~7.   All the details on the cluster model and active space construction as well as on the (vertical) many-body wavefunctions, energies, and relaxation effects are found in Supplementary Note 2. The convergence of all studied quantities with respect to cluster size is given in the Supplementary Tables. 

All ab initio computations were carried out using the ORCA program\cite{Neese_orca6}, version 6.0.1. 
Results in the main text are presented for cc-pVDZ basis set~\cite{cc-pVDZ} using the RIJCOSX~\cite{NEESE2009} density fitting approximation with def2/J~\cite{Weigend2006} auxiliary basis set in the self-consistent field (SCF) calculations and with the cc-pVDZ/C~\cite{Weigend2002} auxiliary basis set in post-SCF calculations, respectively. 

Zero-field splitting parameters were obtained by the quasi-degenerate perturbation theory (QDPT)\cite{Roemelt_2013}, summarized in Supplementary Note~1, as implemented in the ORCA program suite.
The intersystem crossing and photoluminescence transition properties were studied within the 1D effective phonon approximation, as summarized in  Supplementary Note~1. The calculations were performed by our in-house developed code, which takes all the necessary input from the a priori ORCA computations. 


\section*{\large Data availability}

The datasets generated and analyzed during the current study are available in this article and its Supplementary Information. The complete computational output files generated during this study are not publicly available due to their large volume but are available from the corresponding author on reasonable request.

\section*{\large Competing interests}

The authors declare no competing financial or non-financial interests

\section*{\large Acknowledgments} 
The computations were enabled by resources provided by the National Academic Infrastructure for Supercomputing in Sweden (NAISS) and the Swedish National Infrastructure for Computing (SNIC) at NSC, partially funded by the Swedish Research
Council through grant agreements no. 2022-06725.
We acknowledge Digital Government Development and Project Management Ltd. for awarding us access to the Komondor HPC facility based in Hungary

\section*{\large Funding} 

This research was supported by the National Research, Development, and Innovation Office of Hungary within the Quantum Information National Laboratory of Hungary (Grant No. 2022-2.1.1-NL-2022-00004) and within grant FK 145395. We acknowledge the support of the European Union under Horizon Europe for the QRC-4-ESP project (Grant Agreement 101129663) and the QUEST project (Grant Agreement 101156088)
Z.B.  acknowledges the financial support of the János Bolyai Research Fellowship of the Hungarian Academy of Sciences.


\section*{\large References}
\vskip 0.5cm
\bibliography{references_cited_only}{}

@misc{Nakamura2026,
  author        = {Nakamura, Yuki and Iwasaki, Takuya and Nakaharai, Shu and Ogawa, Shinichi and Morita, Yukinori and Watanabe, Kenji and Taniguchi, Takashi and Sasaki, Kento and Kobayashi, Kensuke},
  title         = {Spin-coherence characterization of boron vacancy defects in hexagonal boron nitride with broadband microwave pulses},
  year          = {2026},
  eprint        = {2604.27843},
  archivePrefix = {arXiv},
  primaryClass  = {cond-mat.mes-hall}
}

@article{weber_quantum_2010,
  author  = {Weber, J. R. and Koehl, W. F. and Varley, J. B. and Janotti, A. and Buckley, B. B. and Van de Walle, C. G. and Awschalom, D. D.},
  title   = {Quantum computing with defects},
  journal = {Proceedings of the National Academy of Sciences},
  volume  = {107},
  number  = {19},
  pages   = {8513--8518},
  year    = {2010},
  doi     = {10.1073/pnas.1003052107},
  url     = {https://doi.org/10.1073/pnas.1003052107}
}

@article{wolfowicz_quantum_2021,
  author  = {Wolfowicz, Gary and Heremans, F. Joseph and Anderson, Christopher P. and Kanai, Shun and Seo, Hosung and Gali, Adam and Galli, Giulia and Awschalom, David D.},
  title   = {Quantum guidelines for solid-state spin defects},
  journal = {Nature Reviews Materials},
  volume  = {6},
  number  = {10},
  pages   = {906--925},
  year    = {2021},
  doi     = {10.1038/s41578-021-00306-y},
  url     = {https://doi.org/10.1038/s41578-021-00306-y}
}

@article{Kianinia_2022,
  author  = {Kianinia, Mehran and Xu, Zai-Quan and Toth, Milos and Aharonovich, Igor},
  title   = {Quantum emitters in {2D} materials: Emitter engineering, photophysics, and integration in photonic nanostructures},
  journal = {Applied Physics Reviews},
  volume  = {9},
  number  = {1},
  pages   = {011306},
  year    = {2022},
  doi     = {10.1063/5.0072091},
  url     = {https://doi.org/10.1063/5.0072091}
}

@article{esmann2024,
  author  = {Esmann, Martin and Wein, Stephen C. and Antón-Solanas, Carlos},
  title   = {Solid-State Single-Photon Sources: Recent Advances for Novel Quantum Materials},
  journal = {Advanced Functional Materials},
  volume  = {34},
  number  = {30},
  pages   = {2315936},
  year    = {2024},
  doi     = {10.1002/adfm.202315936},
  url     = {https://doi.org/10.1002/adfm.202315936}
}

@article{Doherty2011,
  author  = {M W Doherty and N B Manson and P Delaney and L C L Hollenberg},
  title   = {The negatively charged nitrogen-vacancy centre in diamond: the electronic solution},
  journal = {New Journal of Physics},
  volume  = {13},
  number  = {2},
  pages   = {025019},
  year    = {2011},
  doi     = {10.1088/1367-2630/13/2/025019},
  url     = {https://doi.org/10.1088/1367-2630/13/2/025019}
}

@article{rondin_magnetometry_2014,
  author  = {Rondin, L and Tetienne, J-P and Hingant, T and Roch, J-F and Maletinsky, P and Jacques, V},
  title   = {Magnetometry with nitrogen-vacancy defects in diamond},
  journal = {Reports on Progress in Physics},
  volume  = {77},
  number  = {5},
  pages   = {056503},
  year    = {2014},
  doi     = {10.1088/0034-4885/77/5/056503},
  url     = {https://doi.org/10.1088/0034-4885/77/5/056503}
}

@article{rovny_nanoscale_2024,
  author  = {Rovny, Jared and Gopalakrishnan, Sarang and Jayich, Ania C. Bleszynski and Maletinsky, Patrick and Demler, Eugene and de Leon, Nathalie P.},
  title   = {Nanoscale diamond quantum sensors for many-body physics},
  journal = {Nature Reviews Physics},
  volume  = {6},
  number  = {12},
  pages   = {753--768},
  year    = {2024},
  doi     = {10.1038/s42254-024-00775-4},
  url     = {https://doi.org/10.1038/s42254-024-00775-4}
}

@article{romach_spectroscopy_2015,
  author  = {Romach, Y. and Müller, C. and Unden, T. and Rogers, L.J. and Isoda, T. and Itoh, K.M. and Markham, M. and Stacey, A. and Meijer, J. and Pezzagna, S. and Naydenov, B. and {McGuinness}, L.P. and Bar-Gill, N. and Jelezko, F.},
  title   = {Spectroscopy of Surface-Induced Noise Using Shallow Spins in Diamond},
  journal = {Physical Review Letters},
  volume  = {114},
  number  = {1},
  pages   = {017601},
  year    = {2015},
  doi     = {10.1103/PhysRevLett.114.017601},
  url     = {https://doi.org/10.1103/PhysRevLett.114.017601}
}

@article{sangtawesin_origins_2019,
  author  = {Sangtawesin, Sorawis and Dwyer, Bo L. and Srinivasan, Srikanth and Allred, James J. and Rodgers, Lila V.H. and De Greve, Kristiaan and Stacey, Alastair and Dontschuk, Nikolai and O’Donnell, Kane M. and Hu, Di and Evans, D. Andrew and Jaye, Cherno and Fischer, Daniel A. and Markham, Matthew L. and Twitchen, Daniel J. and Park, Hongkun and Lukin, Mikhail D. and de Leon, Nathalie P.},
  title   = {Origins of Diamond Surface Noise Probed by Correlating Single-Spin Measurements with Surface Spectroscopy},
  journal = {Physical Review X},
  volume  = {9},
  number  = {3},
  pages   = {031052},
  year    = {2019},
  doi     = {10.1103/PhysRevX.9.031052},
  url     = {https://doi.org/10.1103/PhysRevX.9.031052}
}

@article{tran_quantum_2016,
  author  = {Tran, Toan Trong and Bray, Kerem and Ford, Michael J. and Toth, Milos and Aharonovich, Igor},
  title   = {Quantum emission from hexagonal boron nitride monolayers},
  journal = {Nature Nanotechnology},
  volume  = {11},
  number  = {1},
  pages   = {37--41},
  year    = {2016},
  doi     = {10.1038/nnano.2015.242},
  url     = {https://doi.org/10.1038/nnano.2015.242}
}

@article{gottscholl_2020,
  author  = {Andreas Gottscholl and Matthias Diez and Victor Soltamov and Christian Kasper and Andreas Sperlich and Mehran Kianinia and Carlo Bradac and Igor Aharonovich and Vladimir Dyakonov},
  title   = {Room temperature coherent control of spin defects in hexagonal boron nitride},
  journal = {Science Advances},
  volume  = {7},
  number  = {14},
  pages   = {eabf3630},
  year    = {2021},
  doi     = {10.1126/sciadv.abf3630},
  url     = {https://doi.org/10.1126/sciadv.abf3630}
}

@article{chejanovsky_single-spin_2021,
  author  = {Chejanovsky, Nathan and Mukherjee, Amlan and Geng, Jianpei and Chen, Yu-Chen and Kim, Youngwook and Denisenko, Andrej and Finkler, Amit and Taniguchi, Takashi and Watanabe, Kenji and Dasari, Durga Bhaktavatsala Rao and Auburger, Philipp and Gali, Adam and Smet, Jurgen H. and Wrachtrup, Jörg},
  title   = {Single-spin resonance in a van der {Waals} embedded paramagnetic defect},
  journal = {Nature Materials},
  volume  = {20},
  number  = {8},
  pages   = {1079--1084},
  year    = {2021},
  doi     = {10.1038/s41563-021-00979-4},
  url     = {https://doi.org/10.1038/s41563-021-00979-4}
}

@article{stern_room-temperature_2022,
  author  = {Stern, Hannah L. and Gu, Qiushi and Jarman, John and Eizagirre Barker, Simone and Mendelson, Noah and Chugh, Dipankar and Schott, Sam and Tan, Hoe H. and Sirringhaus, Henning and Aharonovich, Igor and Atatüre, Mete},
  title   = {Room-temperature optically detected magnetic resonance of single defects in hexagonal boron nitride},
  journal = {Nature Communications},
  volume  = {13},
  number  = {1},
  pages   = {618},
  year    = {2022},
  doi     = {10.1038/s41467-022-28169-z},
  url     = {https://doi.org/10.1038/s41467-022-28169-z}
}

@article{gao_single_2025,
  author  = {Gao, Xingyu and Vaidya, Sumukh and Li, Kejun and Ge, Zhun and Dikshit, Saakshi and Zhang, Shimin and Ju, Peng and Shen, Kunhong and Jin, Yuanbin and Ping, Yuan and Li, Tongcang},
  title   = {Single nuclear spin detection and control in a van der {Waals} material},
  journal = {Nature},
  volume  = {643},
  number  = {8073},
  pages   = {943--949},
  year    = {2025},
  doi     = {10.1038/s41586-025-09258-7},
  url     = {https://doi.org/10.1038/s41586-025-09258-7}
}

@article{robertson_charge_2025,
  author  = {Robertson, Islay O. and Whitefield, Benjamin and Scholten, Sam C. and Singh, Priya and Healey, Alexander J. and Reineck, Philipp and Kianinia, Mehran and Barcza, Gergely and Ivády, Viktor and Broadway, David A. and Aharonovich, Igor and Tetienne, Jean-Philippe},
  title   = {A charge transfer mechanism for optically addressable solid-state spin pairs},
  journal = {Nature Physics},
  volume  = {21},
  number  = {12},
  pages   = {1981--1987},
  year    = {2025},
  doi     = {10.1038/s41567-025-03091-5},
  url     = {https://doi.org/10.1038/s41567-025-03091-5}
}

@article{whitefield_narrowband_2026,
  author  = {Whitefield, Benjamin and Zeng, Helen Zhi Jie and Liddle-Wesolowski, James and Robertson, Islay O. and Ganyecz, Ádám and Ivády, Viktor and Watanabe, Kenji and Taniguchi, Takashi and Toth, Milos and Tetienne, Jean-Philippe and Aharonovich, Igor and Kianinia, Mehran},
  title   = {Narrowband quantum emitters in hexagonal boron nitride with optically addressable spins},
  journal = {Nature Materials},
  volume  = {25},
  number  = {3},
  pages   = {412--419},
  year    = {2026},
  doi     = {10.1038/s41563-025-02458-6},
  url     = {https://doi.org/10.1038/s41563-025-02458-6}
}

@article{tetienne_quantum_2021,
  author  = {Tetienne, J.-P.},
  title   = {Quantum sensors go flat},
  journal = {Nature Physics},
  volume  = {17},
  number  = {10},
  pages   = {1074--1075},
  year    = {2021},
  doi     = {10.1038/s41567-021-01338-5},
  url     = {https://doi.org/10.1038/s41567-021-01338-5}
}

@article{healey_quantum_2022,
  author  = {Healey, A. J. and Scholten, S. C. and Yang, T. and Scott, J. A. and Abrahams, G. J. and Robertson, I. O. and Hou, X. F. and Guo, Y. F. and Rahman, S. and Lu, Y. and Kianinia, M. and Aharonovich, I. and Tetienne, J.-P.},
  title   = {Quantum microscopy with van der {Waals} heterostructures},
  journal = {Nature Physics},
  volume  = {19},
  number  = {1},
  pages   = {87--91},
  year    = {2023},
  doi     = {10.1038/s41567-022-01815-5},
  url     = {https://doi.org/10.1038/s41567-022-01815-5}
}

@article{ivady_ab_2020,
  author  = {Ivády, Viktor and Barcza, Gergely and Thiering, Gergő and Li, Song and Hamdi, Hanen and Chou, Jyh-Pin and Legeza, \"{O}rs and Gali, Adam},
  title   = {Ab initio theory of the negatively charged boron vacancy qubit in hexagonal boron nitride},
  journal = {npj Computational Materials},
  volume  = {6},
  number  = {1},
  pages   = {41},
  year    = {2020},
  doi     = {10.1038/s41524-020-0305-x},
  url     = {https://doi.org/10.1038/s41524-020-0305-x}
}

@article{gottscholl_initialization_2020,
  author  = {Gottscholl, Andreas and Kianinia, Mehran and Soltamov, Victor and Orlinskii, Sergei and Mamin, Georgy and Bradac, Carlo and Kasper, Christian and Krambrock, Klaus and Sperlich, Andreas and Toth, Milos and Aharonovich, Igor and Dyakonov, Vladimir},
  title   = {Initialization and read-out of intrinsic spin defects in a van der {Waals} crystal at room temperature},
  journal = {Nature Materials},
  volume  = {19},
  number  = {5},
  pages   = {540--545},
  year    = {2020},
  doi     = {10.1038/s41563-020-0619-6},
  url     = {https://doi.org/10.1038/s41563-020-0619-6}
}

@article{gottscholl_spin_2021,
  author  = {Gottscholl, Andreas and Diez, Matthias and Soltamov, Victor and Kasper, Christian and Krauße, Dominik and Sperlich, Andreas and Kianinia, Mehran and Bradac, Carlo and Aharonovich, Igor and Dyakonov, Vladimir},
  title   = {Spin defects in {hBN} as promising temperature, pressure and magnetic field quantum sensors},
  journal = {Nature Communications},
  volume  = {12},
  number  = {1},
  pages   = {4480},
  year    = {2021},
  doi     = {10.1038/s41467-021-24725-1},
  url     = {https://doi.org/10.1038/s41467-021-24725-1}
}

@article{gao_high-contrast_2021,
  author  = {Gao, Xingyu and Jiang, Boyang and Llacsahuanga Allcca, Andres E. and Shen, Kunhong and Sadi, Mohammad A. and Solanki, Abhishek B. and Ju, Peng and Xu, Zhujing and Upadhyaya, Pramey and Chen, Yong P. and Bhave, Sunil A. and Li, Tongcang},
  title   = {High-Contrast Plasmonic-Enhanced Shallow Spin Defects in Hexagonal Boron Nitride for Quantum Sensing},
  journal = {Nano Letters},
  volume  = {21},
  number  = {18},
  pages   = {7708--7714},
  year    = {2021},
  doi     = {10.1021/acs.nanolett.1c02495},
  url     = {https://doi.org/10.1021/acs.nanolett.1c02495}
}

@article{lyu_strain_2022,
  author  = {Lyu, Xiaodan and Tan, Qinghai and Wu, Lishu and Zhang, Chusheng and Zhang, Zhaowei and Mu, Zhao and Zúñiga-Pérez, Jesús and Cai, Hongbing and Gao, Weibo},
  title   = {Strain Quantum Sensing with Spin Defects in Hexagonal Boron Nitride},
  journal = {Nano Letters},
  volume  = {22},
  number  = {16},
  pages   = {6553--6559},
  year    = {2022},
  doi     = {10.1021/acs.nanolett.2c01722},
  url     = {https://doi.org/10.1021/acs.nanolett.2c01722}
}

@article{daly_prospects_2025,
  author  = {Daly, Declan M. and Reed, Niko R. and {DeVience}, Stephen J. and Yin, Zechuan and Cremer, Johannes and Beling, Andrew J. and Blanchard, John W. and Walsworth, Ronald L.},
  title   = {Prospects for Ultralow-Mass Nuclear Magnetic Resonance using Spin Defects in Hexagonal Boron Nitride},
  journal = {arXiv:2505.00383},
  year    = {2025},
  doi     = {10.48550/arXiv.2505.00383},
  url     = {https://doi.org/10.48550/arXiv.2505.00383}
}

@article{biswas_quantum_2025,
  author  = {Biswas, Souvik and Scuri, Giovanni and Huffman, Noah and Rosenthal, Eric I. and Gong, Ruotian and Poirier, Thomas and Gao, Xingyu and Vaidya, Sumukh and Stein, Abigail J. and Weissman, Tsachy and Edgar, James H. and Li, Tongcang and Zu, Chong and Vučković, Jelena and Choi, Joonhee},
  title   = {Quantum sensing with a spin ensemble in a two-dimensional material},
  journal = {arXiv:2509.08984},
  year    = {2025},
  doi     = {10.48550/arXiv.2509.08984},
  url     = {https://doi.org/10.48550/arXiv.2509.08984}
}

@article{Attaccalite2011,
  author  = {Attaccalite, C. and Bockstedte, M. and Marini, A. and Rubio, A. and Wirtz, L.},
  title   = {Coupling of excitons and defect states in boron-nitride nanostructures},
  journal = {Physical Review B},
  volume  = {83},
  number  = {14},
  pages   = {144115},
  year    = {2011},
  doi     = {10.1103/PhysRevB.83.144115},
  url     = {https://doi.org/10.1103/PhysRevB.83.144115}
}

@article{Tawfik2017,
  author  = {Tawfik, Sherif Abdulkader and Ali, Sajid and Fronzi, Marco and Kianinia, Mehran and Tran, Toan Trong and Stampfl, Catherine and Aharonovich, Igor and Toth, Milos and Ford, Michael J.},
  title   = {First-principles investigation of quantum emission from {hBN} defects},
  journal = {Nanoscale},
  volume  = {9},
  number  = {36},
  pages   = {13575--13582},
  year    = {2017},
  doi     = {10.1039/C7NR04270A},
  url     = {https://doi.org/10.1039/C7NR04270A}
}

@article{Abdi2018,
  author  = {Abdi, Mehdi and Chou, Jyh-Pin and Gali, Adam and Plenio, Martin B.},
  title   = {Color Centers in Hexagonal Boron Nitride Monolayers: A Group Theory and Ab Initio Analysis},
  journal = {ACS Photonics},
  volume  = {5},
  number  = {5},
  pages   = {1967--1976},
  year    = {2018},
  doi     = {10.1021/acsphotonics.7b01442},
  url     = {https://doi.org/10.1021/acsphotonics.7b01442}
}

@article{weston_native_2018,
  author  = {Weston, L. and Wickramaratne, D. and Mackoit, M. and Alkauskas, A. and Van de Walle, C. G.},
  title   = {Native point defects and impurities in hexagonal boron nitride},
  journal = {Physical Review B},
  volume  = {97},
  number  = {21},
  pages   = {214104},
  year    = {2018},
  doi     = {10.1103/PhysRevB.97.214104},
  url     = {https://doi.org/10.1103/PhysRevB.97.214104}
}

@article{sajid_edge_2020,
  author  = {Sajid, A. and Thygesen, Kristian S. and Reimers, Jeffrey R. and Ford, Michael J.},
  title   = {Edge effects on optically detected magnetic resonance of vacancy defects in hexagonal boron nitride},
  journal = {Communications Physics},
  volume  = {3},
  number  = {1},
  pages   = {153},
  year    = {2020},
  doi     = {10.1038/s42005-020-00416-z},
  url     = {https://doi.org/10.1038/s42005-020-00416-z}
}

@article{reimers_photoluminescence_2020,
  author  = {Reimers, Jeffrey R. and Shen, Jun and Kianinia, Mehran and Bradac, Carlo and Aharonovich, Igor and Ford, Michael J. and Piecuch, Piotr},
  title   = {Photoluminescence, photophysics, and photochemistry of the {V}{$_{\mathrm{B}}^{-}$} defect in hexagonal boron nitride},
  journal = {Physical Review B},
  volume  = {102},
  number  = {14},
  pages   = {144105},
  year    = {2020},
  doi     = {10.1103/PhysRevB.102.144105},
  url     = {https://doi.org/10.1103/PhysRevB.102.144105}
}

@article{Barcza_2021,
  author  = {Barcza, Gergely and Ivády, Viktor and Szilvási, Tibor and V{\"o}r{\"o}s, Márton and Veis, Libor and Gali, Ádám and Legeza, {\"O}rs},
  title   = {{DMRG} on Top of Plane-Wave {Kohn}–{Sham} Orbitals: A Case Study of Defected Boron Nitride},
  journal = {Journal of Chemical Theory and Computation},
  volume  = {17},
  number  = {2},
  pages   = {1143--1154},
  year    = {2021},
  doi     = {10.1021/acs.jctc.0c00809},
  url     = {https://doi.org/10.1021/acs.jctc.0c00809}
}

@article{Reimers2018,
  author  = {Reimers, Jeffrey R. and Sajid, A. and Kobayashi, Rika and Ford, Michael J.},
  title   = {Understanding and Calibrating Density-Functional-Theory Calculations Describing the Energy and Spectroscopy of Defect Sites in Hexagonal Boron Nitride},
  journal = {Journal of Chemical Theory and Computation},
  volume  = {14},
  number  = {3},
  pages   = {1602--1613},
  year    = {2018},
  doi     = {10.1021/acs.jctc.7b01072},
  url     = {https://doi.org/10.1021/acs.jctc.7b01072}
}

@article{benedek_accurate_2025,
  author  = {Benedek, Zsolt and Ganyecz, \'Ad\'am and Pershin, Anton and Ivády, Viktor and Barcza, Gergely},
  title   = {Accurate and convergent energetics of color centers by wavefunction theory},
  journal = {npj Computational Materials},
  volume  = {11},
  number  = {1},
  pages   = {346},
  year    = {2025},
  doi     = {10.1038/s41524-025-01813-0},
  url     = {https://doi.org/10.1038/s41524-025-01813-0}
}

@article{luu_identifying_2025,
  author  = {Luu, Minh Tuan and Linderälv, Christopher and Benedek, Zsolt and Ganyecz, \'Ad\'am and Barcza, Gergely and Ivády, Viktor and Ulbricht, Ronald},
  title   = {Identifying high-energy electronic states of {NV}{$^{-}$} centers in diamond},
  journal = {Applied Physics Letters},
  volume  = {126},
  number  = {23},
  pages   = {234001},
  year    = {2025},
  doi     = {10.1063/5.0268247},
  url     = {https://doi.org/10.1063/5.0268247}
}

@article{Mathur2022,
  author  = {Mathur, Nikhil and Mukherjee, Arunabh and Gao, Xingyu and Luo, Jialun and McCullian, Brendan A. and Li, Tongcang and Vamivakas, A. Nick and Fuchs, Gregory D.},
  title   = {Excited-state spin-resonance spectroscopy of {V}{$_{\mathrm{B}}^{-}$} defect centers in hexagonal boron nitride},
  journal = {Nature Communications},
  volume  = {13},
  number  = {1},
  pages   = {3233},
  year    = {2022},
  doi     = {10.1038/s41467-022-30772-z},
  url     = {https://doi.org/10.1038/s41467-022-30772-z}
}

@article{geng_deformation-driven_2025,
  author  = {Geng, Jianpei and Zhou, Xuankai and Gross, Nils and Li, Song and Kong, Yan Tung and Zhang, Jixing and Bian, Guodong and Ng, San Lam and Ho, Cheng-I and Denisenko, Andrej and St{\"o}hr, Rainer and Peng, Ruoming and Smet, Jurgen and Wrachtrup, J{\"o}rg},
  title   = {Deformation-Driven Enhancement of Spin Defect Emission in Hexagonal Boron Nitride},
  journal = {ACS Nano},
  volume  = {20},
  number  = {4},
  pages   = {3510--3518},
  year    = {2026},
  doi     = {10.1021/acsnano.5c15247},
  url     = {https://doi.org/10.1021/acsnano.5c15247}
}

@article{libbi_phonon-assisted_2022,
  author  = {Libbi, Francesco and de Melo, Pedro Miguel M. C. and Zanolli, Zeila and Verstraete, Matthieu Jean and Marzari, Nicola},
  title   = {Phonon-Assisted Luminescence in Defect Centers from Many-Body Perturbation Theory},
  journal = {Physical Review Letters},
  volume  = {128},
  number  = {16},
  pages   = {167401},
  year    = {2022},
  doi     = {10.1103/PhysRevLett.128.167401},
  url     = {https://doi.org/10.1103/PhysRevLett.128.167401}
}

@article{whitefield_magnetic_2023,
  author  = {Whitefield, Benjamin and Toth, Milos and Aharonovich, Igor and Tetienne, Jean-Philippe and Kianinia, Mehran},
  title   = {Magnetic Field Sensitivity Optimization of Negatively Charged Boron Vacancy Defects in {hBN}},
  journal = {Advanced Quantum Technologies},
  volume  = {8},
  number  = {4},
  pages   = {2300118},
  year    = {2023},
  doi     = {10.1002/qute.202300118},
  url     = {https://doi.org/10.1002/qute.202300118}
}

@article{baber_excited_2022,
  author  = {Baber, Simon and Malein, Ralph Nicholas Edward and Khatri, Prince and Keatley, Paul Steven and Guo, Shi and Withers, Freddie and Ramsay, Andrew J. and Luxmoore, Isaac J.},
  title   = {Excited State Spectroscopy of Boron Vacancy Defects in Hexagonal Boron Nitride Using Time-Resolved Optically Detected Magnetic Resonance},
  journal = {Nano Letters},
  volume  = {22},
  number  = {1},
  pages   = {461--467},
  year    = {2022},
  doi     = {10.1021/acs.nanolett.1c04366},
  url     = {https://doi.org/10.1021/acs.nanolett.1c04366}
}

@article{VB-ZPL,
  author  = {Qian, Chenjiang and Villafa{\~n}e, Viviana and Schalk, Martin and Astakhov, G. V. and Kentsch, Ulrich and Helm, Manfred and Soubelet, Pedro and Wilson, Nathan P. and Rizzato, Roberto and Mohr, Stephan and Holleitner, Alexander W. and Bucher, Dominik B. and Stier, Andreas V. and Finley, Jonathan J.},
  title   = {Unveiling the Zero-Phonon Line of the Boron Vacancy Center by Cavity-Enhanced Emission},
  journal = {Nano Letters},
  volume  = {22},
  number  = {13},
  pages   = {5137--5142},
  year    = {2022},
  doi     = {10.1021/acs.nanolett.2c00739},
  url     = {https://doi.org/10.1021/acs.nanolett.2c00739}
}

@article{babar_low-symmetry_2024,
  author  = {Babar, Rohit and Barcza, Gergely and Pershin, Anton and Park, Hyoju and Bulancea Lindvall, Oscar and Thiering, Gergő and Legeza, Örs and Warner, Jamie H. and Abrikosov, Igor A. and Gali, Adam and Ivády, Viktor},
  title   = {Low-symmetry vacancy-related spin qubit in hexagonal boron nitride},
  journal = {npj Computational Materials},
  volume  = {10},
  number  = {1},
  pages   = {184},
  year    = {2024},
  doi     = {10.1038/s41524-024-01361-z},
  url     = {https://doi.org/10.1038/s41524-024-01361-z}
}

@article{Smart2021,
  author  = {Smart, Tyler J. and Li, Kejun and Xu, Junqing and Ping, Yuan},
  title   = {Intersystem crossing and exciton–defect coupling of spin defects in hexagonal boron nitride},
  journal = {npj Computational Materials},
  volume  = {7},
  number  = {1},
  pages   = {59},
  year    = {2021},
  doi     = {10.1038/s41524-021-00525-5},
  url     = {https://doi.org/10.1038/s41524-021-00525-5}
}

@article{li_carbon_2022,
  author  = {Li, Kejun and Smart, Tyler J. and Ping, Yuan},
  title   = {Carbon trimer as a 2 {eV} single-photon emitter candidate in hexagonal boron nitride: A first-principles study},
  journal = {Physical Review Materials},
  volume  = {6},
  number  = {4},
  pages   = {L042201},
  year    = {2022},
  doi     = {10.1103/PhysRevMaterials.6.L042201},
  url     = {https://doi.org/10.1103/PhysRevMaterials.6.L042201}
}

@article{PhysRevB.110.184302,
  author  = {Li, Kejun and Dergachev, Vsevolod D. and Dergachev, Ilya D. and Zhang, Shimin and Varganov, Sergey A. and Ping, Yuan},
  title   = {Excited-state dynamics and optically detected magnetic resonance of solid-state spin defects from first principles},
  journal = {Physical Review B},
  volume  = {110},
  number  = {18},
  pages   = {184302},
  year    = {2024},
  doi     = {10.1103/PhysRevB.110.184302},
  url     = {https://doi.org/10.1103/PhysRevB.110.184302}
}

@article{Jin2025ISC,
  author  = {Jin, Yu and Park, Jinsoo and McMillan, Marquis M. and Ohm, Daniel Donghyon and Barnes, Corrie and Pingault, Benjamin and Egerstrom, Christopher and Huang, Benchen and Govoni, Marco and Heremans, F. Joseph and Awschalom, David D. and Galli, Giulia},
  title   = {First-Principles Framework for the Prediction of Intersystem Crossing Rates in Spin Defects: The Role of Electron Correlation},
  journal = {Physical Review Letters},
  volume  = {135},
  pages   = {036401},
  year    = {2025},
  doi     = {10.1103/nw3r-zy8q},
  url     = {https://doi.org/10.1103/nw3r-zy8q}
}

@article{clua-provost_spin-dependent_2024,
  author  = {Clua-Provost, T. and Mu, Z. and Durand, A. and Schrader, C. and Happacher, J. and Bocquel, J. and Maletinsky, P. and Fraunié, J. and Marie, X. and Robert, C. and Seine, G. and Janzen, E. and Edgar, J. H. and Gil, B. and Cassabois, G. and Jacques, V.},
  title   = {Spin-dependent photodynamics of boron-vacancy centers in hexagonal boron nitride},
  journal = {Physical Review B},
  volume  = {110},
  number  = {1},
  pages   = {014104},
  year    = {2024},
  doi     = {10.1103/PhysRevB.110.014104},
  url     = {https://doi.org/10.1103/PhysRevB.110.014104}
}

@article{Durand-2023,
  author  = {Durand, A. and Clua-Provost, T. and Fabre, F. and Kumar, P. and Li, J. and Edgar, J. H. and Udvarhelyi, P. and Gali, A. and Marie, X. and Robert, C. and G\'erard, J. M. and Gil, B. and Cassabois, G. and Jacques, V.},
  title   = {Optically Active Spin Defects in Few-Layer Thick Hexagonal Boron Nitride},
  journal = {Physical Review Letters},
  volume  = {131},
  number  = {11},
  pages   = {116902},
  year    = {2023},
  doi     = {10.1103/PhysRevLett.131.116902},
  url     = {https://doi.org/10.1103/PhysRevLett.131.116902}
}

@article{Carbone-2025,
  author  = {Carbone, Amedeo and Breev, Ilia D. and Figueiredo, Johannes and Kretschmer, Silvan and Geilen, Leonard and Ben Mhenni, Amine and Arceri, Johannes and Krasheninnikov, Arkady V. and Wubs, Martijn and Holleitner, Alexander W. and Huck, Alexander and Kastl, Christoph and Stenger, Nicolas},
  title   = {Quantifying the creation of negatively charged boron vacancies in {He}-ion irradiated hexagonal boron nitride},
  journal = {Physical Review Materials},
  volume  = {9},
  number  = {5},
  pages   = {056203},
  year    = {2025},
  doi     = {10.1103/PhysRevMaterials.9.056203},
  url     = {https://doi.org/10.1103/PhysRevMaterials.9.056203}
}

@article{Udvarhelyi-2023,
  author  = {Udvarhelyi, Péter and Clua-Provost, Tristan and Durand, Alrik and Li, Jiahan and Edgar, James H. and Gil, Bernard and Cassabois, Guillaume and Jacques, Vincent and Gali, Adam},
  title   = {A planar defect spin sensor in a two-dimensional material susceptible to strain and electric fields},
  journal = {npj Computational Materials},
  volume  = {9},
  number  = {1},
  pages   = {150},
  year    = {2023},
  doi     = {10.1038/s41524-023-01111-7},
  url     = {https://doi.org/10.1038/s41524-023-01111-7}
}

@article{mu2025magneticimaginghighpressure,
  author  = {Mu, Z. and Fraunié, J. and Durand, A. and Clément, S. and Finco, A. and Rouquette, J. and Hadj-Azzem, A. and Rougemaille, N. and Coraux, J. and Li, J. and Poirier, T. and Edgar, J. H. and Gerber, I. C. and Marie, X. and Gil, B. and Cassabois, G. and Robert, C. and Jacques, V.},
  title   = {Magnetic imaging under high pressure with a spin-based quantum sensor integrated in a van der {Waals} heterostructure},
  journal = {Nature Communications},
  volume  = {16},
  number  = {1},
  pages   = {8574},
  year    = {2025},
  doi     = {10.1038/s41467-025-63580-2},
  url     = {https://doi.org/10.1038/s41467-025-63580-2}
}

@article{Bosak-2006,
  author  = {Bosak, Alexey and Serrano, Jorge and Krisch, Michael and Watanabe, Kenji and Taniguchi, Takashi and Kanda, Hisao},
  title   = {Elasticity of hexagonal boron nitride: Inelastic x-ray scattering measurements},
  journal = {Physical Review B},
  volume  = {73},
  number  = {4},
  pages   = {041402},
  year    = {2006},
  doi     = {10.1103/PhysRevB.73.041402},
  url     = {https://doi.org/10.1103/PhysRevB.73.041402}
}

@incollection{Roos1987,
  author    = {Roos, B. O.},
  title     = {The Complete Active Space Self-Consistent Field Method And Its Applications in Electronic Structure Calculations},
  booktitle = {Advances in Chemical Physics},
  editor    = {Lawley, K. P.},
  publisher = {John Wiley and Sons, Ltd},
  address   = {New York},
  volume    = {69},
  pages     = {399--445},
  year      = {1987},
  doi       = {10.1002/9780470142943.ch7},
  url       = {https://doi.org/10.1002/9780470142943.ch7}
}

@article{Olsen-2011,
  author  = {Olsen, Jeppe},
  title   = {The {CASSCF} method: A perspective and commentary},
  journal = {International Journal of Quantum Chemistry},
  volume  = {111},
  number  = {13},
  pages   = {3267--3272},
  year    = {2011},
  doi     = {10.1002/qua.23107},
  url     = {https://doi.org/10.1002/qua.23107}
}

@article{casscf,
  author  = {Kollmar, Christian and Sivalingam, Kantharuban and Helmich-Paris, Benjamin and Angeli, Celestino and Neese, Frank},
  title   = {A perturbation-based super-{CI} approach for the orbital optimization of a {CASSCF} wave function},
  journal = {Journal of Computational Chemistry},
  volume  = {40},
  number  = {14},
  pages   = {1463--1470},
  year    = {2019},
  doi     = {10.1002/jcc.25801},
  url     = {https://doi.org/10.1002/jcc.25801}
}

@article{Angeli-2001a,
  author  = {Angeli, Celestino and Cimiraglia, Renzo and Malrieu, Jean-Paul},
  title   = {n-electron valence state perturbation theory: A spinless formulation and an efficient implementation of the strongly contracted and of the partially contracted variants},
  journal = {The Journal of Chemical Physics},
  volume  = {117},
  number  = {20},
  pages   = {9138--9153},
  year    = {2002},
  doi     = {10.1063/1.1515317},
  url     = {https://doi.org/10.1063/1.1515317}
}

@article{Angeli-2001,
  author  = {Angeli, C. and Cimiraglia, R. and Evangelisti, S. and Leininger, T. and Malrieu, J.-P.},
  title   = {Introduction of n-electron valence states for multireference perturbation theory},
  journal = {The Journal of Chemical Physics},
  volume  = {114},
  number  = {23},
  pages   = {10252--10264},
  year    = {2001},
  doi     = {10.1063/1.1361246},
  url     = {https://doi.org/10.1063/1.1361246}
}

@article{Angeli2007,
  author  = {Angeli, Celestino and Pastore, Mariachiara and Cimiraglia, Renzo},
  title   = {New perspectives in multireference perturbation theory: the n-electron valence state approach},
  journal = {Theoretical Chemistry Accounts},
  volume  = {117},
  number  = {5},
  pages   = {743--754},
  year    = {2007},
  doi     = {10.1007/s00214-006-0207-0},
  url     = {https://doi.org/10.1007/s00214-006-0207-0}
}

@article{Guo_2021,
  author  = {Guo, Yang and Sivalingam, Kantharuban and Neese, Frank},
  title   = {Approximations of density matrices in {N}-electron valence state second-order perturbation theory ({NEVPT2}). {I}. {Revisiting} the {NEVPT2} construction},
  journal = {The Journal of Chemical Physics},
  volume  = {154},
  number  = {21},
  pages   = {214111},
  year    = {2021},
  doi     = {10.1063/5.0051211},
  url     = {https://doi.org/10.1063/5.0051211}
}

@article{Kollmar_2021,
  author  = {Kollmar, Christian and Sivalingam, Kantharuban and Guo, Yang and Neese, Frank},
  title   = {An efficient implementation of the {NEVPT2} and {CASPT2} methods avoiding higher-order density matrices},
  journal = {The Journal of Chemical Physics},
  volume  = {155},
  number  = {23},
  pages   = {234104},
  year    = {2021},
  doi     = {10.1063/5.0072129},
  url     = {https://doi.org/10.1063/5.0072129}
}

@article{Neese_orca6,
  author  = {Neese, Frank},
  title   = {Software Update: The {ORCA} Program System—Version 6.0},
  journal = {WIREs Computational Molecular Science},
  volume  = {15},
  number  = {2},
  pages   = {e70019},
  year    = {2025},
  doi     = {10.1002/wcms.70019},
  url     = {https://doi.org/10.1002/wcms.70019}
}

@article{cc-pVDZ,
  author  = {Dunning, Thom H.},
  title   = {{Gaussian} basis sets for use in correlated molecular calculations. {I}. {The} atoms boron through neon and hydrogen},
  journal = {The Journal of Chemical Physics},
  volume  = {90},
  number  = {2},
  pages   = {1007--1023},
  year    = {1989},
  doi     = {10.1063/1.456153},
  url     = {https://doi.org/10.1063/1.456153}
}

@article{NEESE2009,
  author  = {Frank Neese and Frank Wennmohs and Andreas Hansen and Ute Becker},
  title   = {Efficient, approximate and parallel {Hartree}–{Fock} and hybrid {DFT} calculations. {A} ‘chain-of-spheres’ algorithm for the {Hartree}–{Fock} exchange},
  journal = {Chemical Physics},
  volume  = {356},
  number  = {1},
  pages   = {98--109},
  year    = {2009},
  doi     = {10.1016/j.chemphys.2008.10.036},
  url     = {https://doi.org/10.1016/j.chemphys.2008.10.036}
}

@article{Weigend2006,
  author  = {Weigend, Florian},
  title   = {Accurate {Coulomb}-fitting basis sets for {H} to {Rn}},
  journal = {Physical Chemistry Chemical Physics},
  volume  = {8},
  number  = {9},
  pages   = {1057--1065},
  year    = {2006},
  doi     = {10.1039/B515623H},
  url     = {https://doi.org/10.1039/B515623H}
}

@article{Weigend2002,
  author  = {Weigend, Florian and Köhn, Andreas and Hättig, Christof},
  title   = {Efficient use of the correlation consistent basis sets in resolution of the identity {MP2} calculations},
  journal = {The Journal of Chemical Physics},
  volume  = {116},
  number  = {8},
  pages   = {3175--3183},
  year    = {2002},
  doi     = {10.1063/1.1445115},
  url     = {https://doi.org/10.1063/1.1445115}
}

@article{Roemelt_2013,
  author  = {Roemelt, Michael and Maganas, Dimitrios and DeBeer, Serena and Neese, Frank},
  title   = {A combined {DFT} and restricted open-shell configuration interaction method including spin-orbit coupling: Application to transition metal {L}-edge {X}-ray absorption spectroscopy},
  journal = {The Journal of Chemical Physics},
  volume  = {138},
  number  = {20},
  pages   = {204101},
  year    = {2013},
  doi     = {10.1063/1.4804607},
  url     = {https://doi.org/10.1063/1.4804607}
}

@article{Lee2025,
  author  = {Lee, W. and Liu, V. S. and Zhang, Z. and Kim, S.
             and Gong, R. and Du, X. and Pham, K. and Poirier, T.
             and Hao, Z. and Edgar, J. H. and Kim, P. and Zu, C.
             and Davis, E. J. and Yao, N. Y.},
  title   = {Intrinsic High-Fidelity Spin Polarization of Charged
             Vacancies in Hexagonal Boron Nitride},
  journal = {Physical Review Letters},
  volume  = {134},
  number  = {9},
  pages   = {096202},
  year    = {2025},
  doi     = {10.1103/PhysRevLett.134.096202}
}

@article{Konrad2026,
  author  = {Konrad, Paul and Kianinia, Mehran and Spencer, Lesley
             and Sperlich, Andreas and Hein, Lukas
             and Steinicke, Selin and Aharonovich, Igor
             and Dyakonov, Vladimir},
  title   = {Intermediate excited state relaxation dynamics of
             boron vacancy spin defects in hexagonal boron nitride},
  journal = {Science Advances},
  volume  = {12},
  number  = {9},
  pages   = {eaea0109},
  year    = {2026},
  doi     = {10.1126/sciadv.aea0109}
}
\end{document}